A Comprehensive Review of Microwave Imaging for Brain Tumour Detection

Sunit Shantanu Digamber Fulari
*Department of Electronics and Communication, Chandigarh University*
Mohali, India

**Abstract: Antennas are known to radiate and absorb radiation. We can use this property of antennas in microwave imaging. In microwave imaging the dielectric of the surrounding objects is considered. The $S_{11}$ parameters are studied. These parameters are different for cells with tumor and without. This helps us in distinguishing between normal and cancerous cells. So this is a property utilized in brain tumor detection.**

Samsuzzaman et.al in their paper Circular slotted patch with defected grounded monopole patch antenna for microwave-based head imaging applications speak about a new, simple patch antenna for microwave imaging applications. In this work the radiating patch has a circular slot and a tiny triangular cut at the corner on the top right side, this patch antenna is designed and manufactured, with a low cost FR4-epoxy substrate which is having a loss tangent of 0.02 and this is having a dielectric constant of 4.4. In this work the proposed monopole sideband antenna can be used for transmitting and receiving of signals which is having a frequency range of 1.22-3.45 GHz. So the mechanism of this antenna is done in a very simple but interesting way. The antenna incorporates a circular patch in the centre which is radiating in nature as shown in the below figure.

Figure 1:

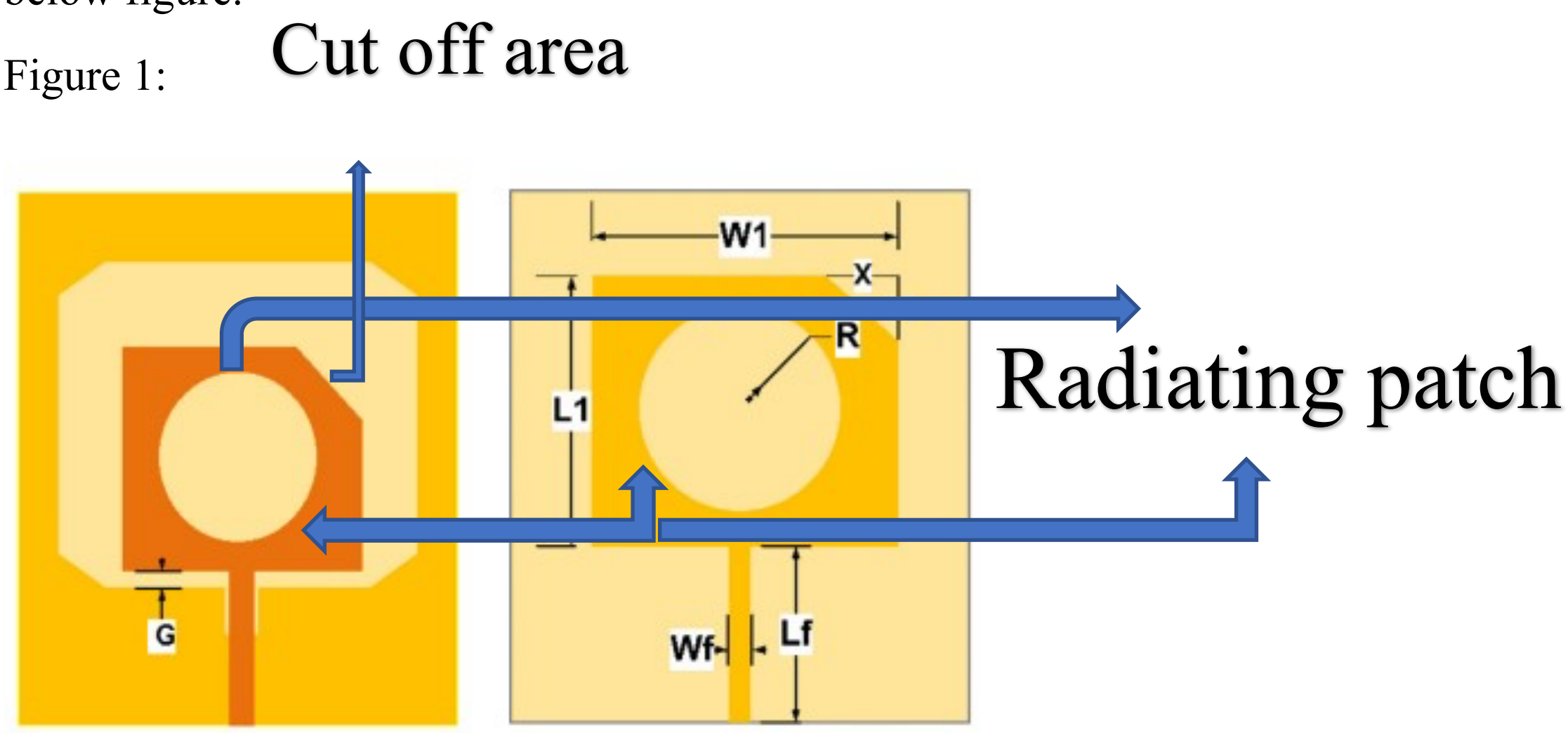

The power is delivered to the device in a very nasty fashion. The 50 ohm impedance microstrip line supplies power to the antenna. In the above figure the top metal is orange, while the bottom metal is yellow and the dielectric substrate is cream coloured. Match is improved by having a notch in the feed at the top right corner of the antenna. In this design the radiating patch shape is squared with a circle cut out at the centre. This is perfectly done in a intelligent manner to reduce the resonance size which is the consequence at the implemented frequency. There is a interesting fact why there is a cut off area in the top right corner of the patch. The cut off corner results in coupling between horizontal and vertical polarized modes. There is also an interesting single polarized mechanism followed in this design. The antenna is 45 degrees polarized and can radiate from both sides of the substrate. When the antenna is kept in a vertical fashion there is a radiation in both east and west direction with deviating angles in all directions unlike in mechanisms when there is only single polarization antenna or radiation in only one direction. When the lower operating frequency is considered, the size keeping in view the electrical properties of the antenna is 0.28λX0.22λX0.005λ( where λ is the lowest wavelength which is the lowest frequency of 1.22Ghz).

$$W_p = \frac{c}{2fr} \int \frac{2}{er+1} \quad (1)$$

$$\mu_{reff} = \frac{\mu r+1}{2} + \frac{\mu r-1}{2} + [1+12\frac{h}{W}]^{-1/2} \quad (2)$$

$$L_p = \frac{c}{2fr\sqrt{\mu reff}} \quad (3)$$

The significance of each of these terms are very important. In the above equation f, c, er, r and h are the resonance frequency, light velocity, material permittivity and substrate height. $\varepsilon_r$ is referring to the effective dielectric constant. In biomedical applications for a good design of antenna, the antenna used is small and compact in size. Mostly to make the design small and compact we use metamaterial antenna. Biomedical applications antenna use compact sized metamaterial material antenna for improved compact size functioning. They performed trial and error parametric study to finalize the final design of the antenna patch and ground design. From the calculations the finalized size of the patch width $W_p$ and patch length $L_p$, 30mm and 30.8mm respectively. This is calculated from the formula above. In antenna design there is a fundamental phenomenon studied fact fully that the size of the antenna or the wavelength of the antenna is less when the frequency is more. As frequency and wavelength are inversely proportional to each other in a relation with velocity. In this paper the design process started with a FR-4 substrate which was 1.5mm in length, there was a conventional traditional rectangular patch used with the ground, this was the integrated initial rudimentary design.

Figure 2:

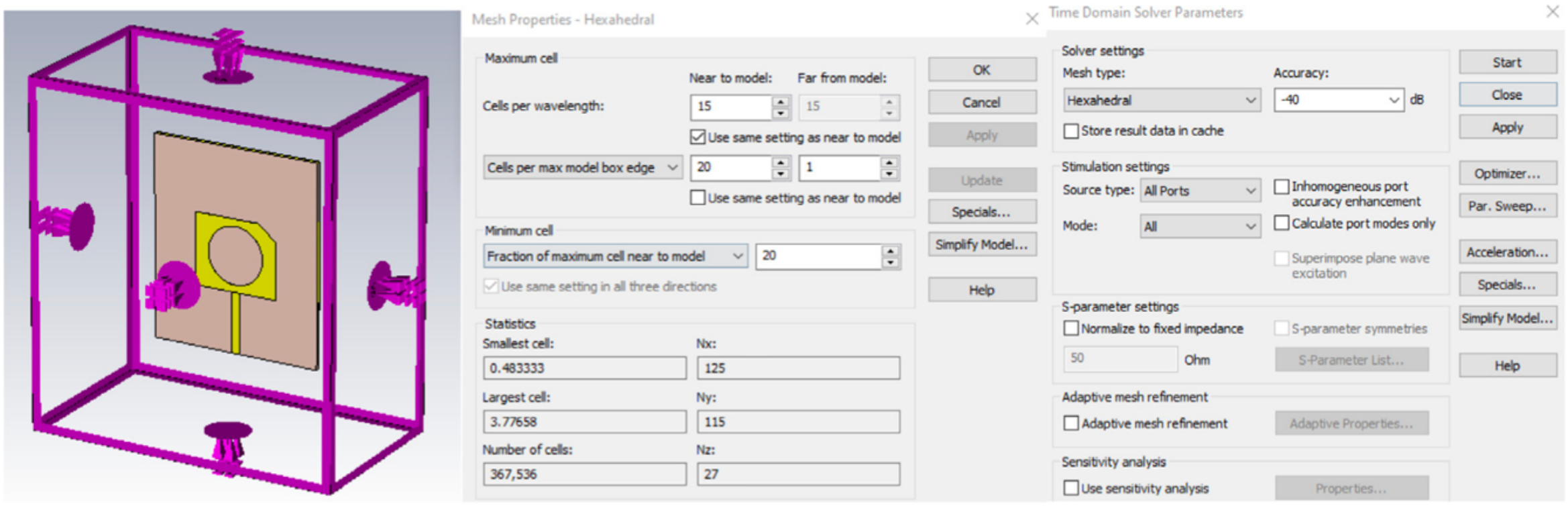

(a) (b) (c)

The antenna design is customized by using CST solver. Further optimization is done in the antenna design. Optimization is done to attain the highest possible bandwidth. In medical application the device or antenna device should behave as a transceiver so optimization is very essential. After looking at the formula and different values of design, after performing optimization the final frequency was set to 2.4 GHz and the final values of patch width $W_p$ and patch length $L_p$ was determined to be 30mm and 25mm. High gain, appropriate or proper frequency of operation and high fidelity is the need of the antenna design. High fidelity of the antenna, fidelity is a term given to realistically design of something, if something has high fidelity means it should behave in reality as a microwave biomedical imaging device or antenna. HFSS simulator was used to verify the design. Normally in this kind of research there is PML boundary conditions used. In our literature review in papers later by there is PML boundary condition used, the logic is that of absorbing antenna in connection with PML boundary condition, the antenna should function as absorbing antenna and there should be minimum reflections. In the papers in this research PML boundary condition the antennas absorb electromagnetic, that is they are impedance matched with the surroundings to minimize reflections. Mostly the fractional bandwidth varies between 0 to 2, in this paper the fractional bandwidth of the antenna is 4. The percentage varies between 0% to 200%. The higher the value of the fractional bandwidth the wider is the bandwidth. Normally a wide band antenna has fractional bandwidth percentage greater than 20%. Ultra-wideband antennas have a bandwidth of 50% or greater. CST and HFSS is used in antenna design. PML boundary are implemented to be used as absorbing material which is matched with the frequency of operation of the antenna and has minimum reflections. There is a contrasting phenomenon observed in HFSS, in this there is a integral equation(IE) which is taken as the boundary condition in radiation.  The design of the process is as follows. The designed antenna has high gain, high fidelity and higher bandwidth leading to better efficiency operation of the device. The frequency is kept low mostly from 1GHz to 4GHz, even 1GHz to achieve an image, though 1GHz is not ideal as the achieved image is noisy and impaired in quality. Microwave imaging requires proper signal electromagnetic transmission to obtain the image, this is the consequential thing in microwave image processing. In the first phase a current rectangular patch with complete ground was selected(dimensions), in step two the ground plane was converted to skewed ground plane. The skewed ground plane reduces the energy absorption, this reduced energy absorption reduces the quality factor Q, as Q factor drops the bandwidth increases. In the third step, the radiating patch was cut into a circular patch in the centre, the end outer part was converted into a rectangular shape. One side at the top was slanted cut out as hypotenuse shape at the end. Tiny triangular cut out shape at the radiator.

Figure 3 and 4:

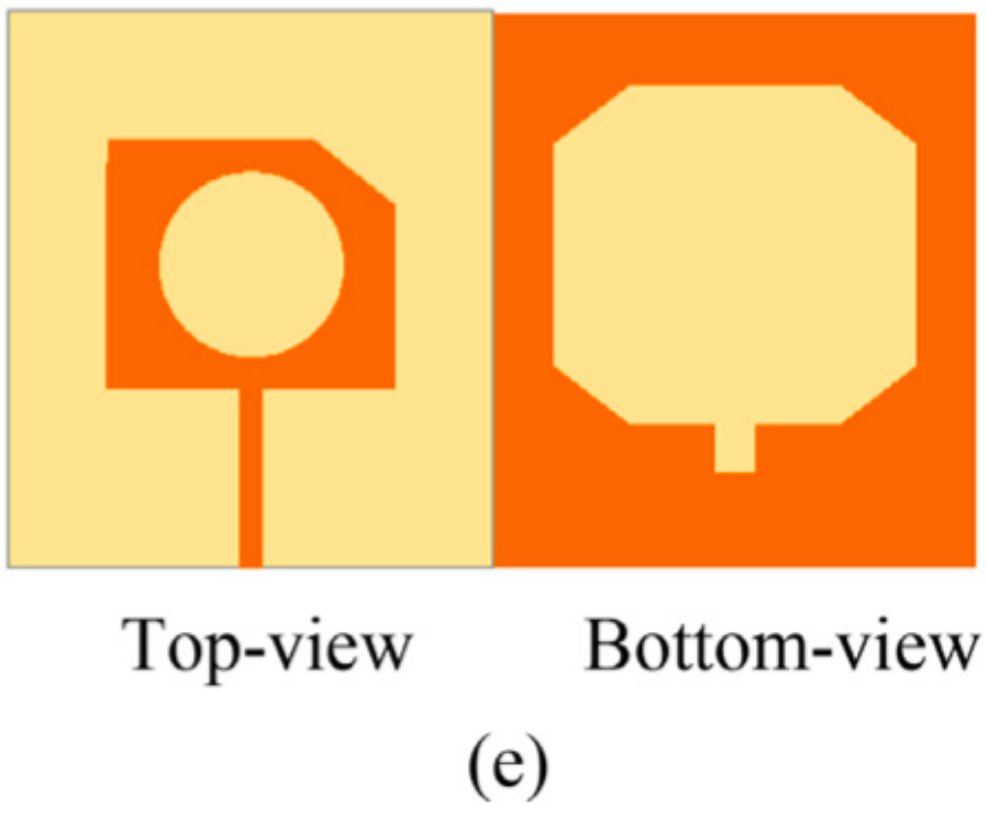

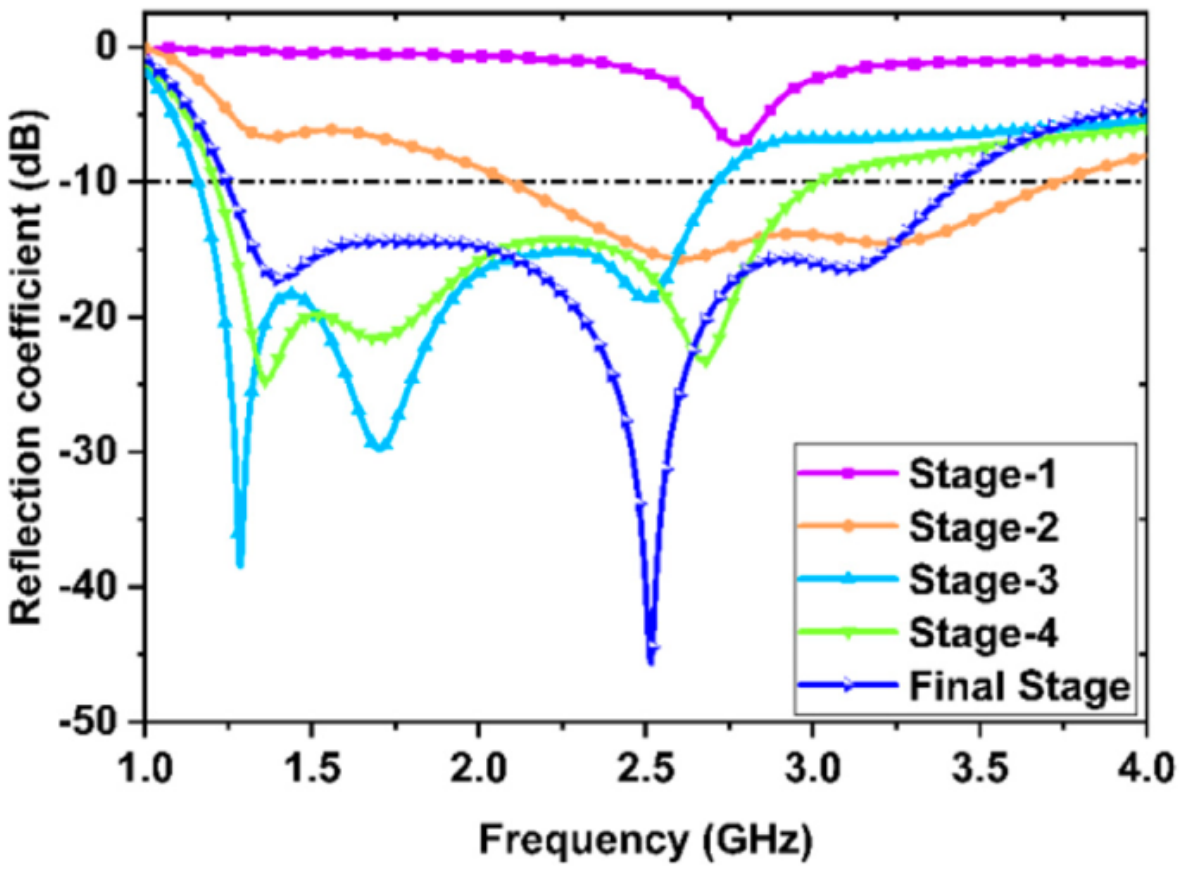

At the feed line there is a rectangular cut out area at the bottom. In the first phase of design (a)

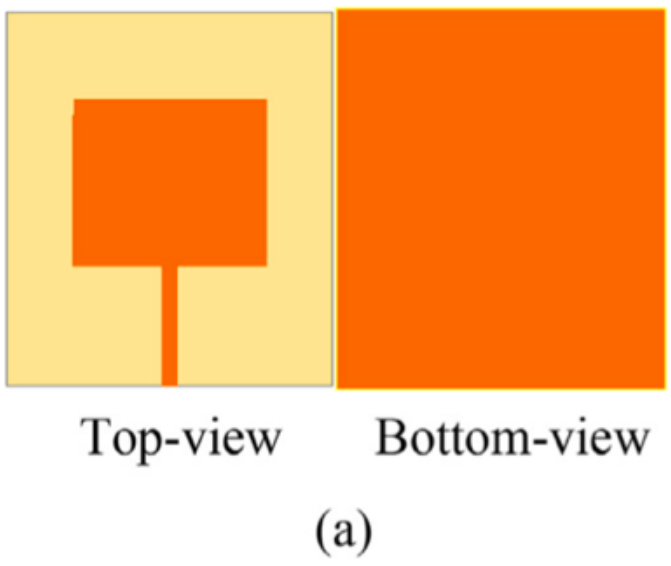

Figure 5: Stage 1.For this design, shown in pink the results could not touch -10 decibel value. This is due to the impedance imbalance between the mediated patch and bottom ground.

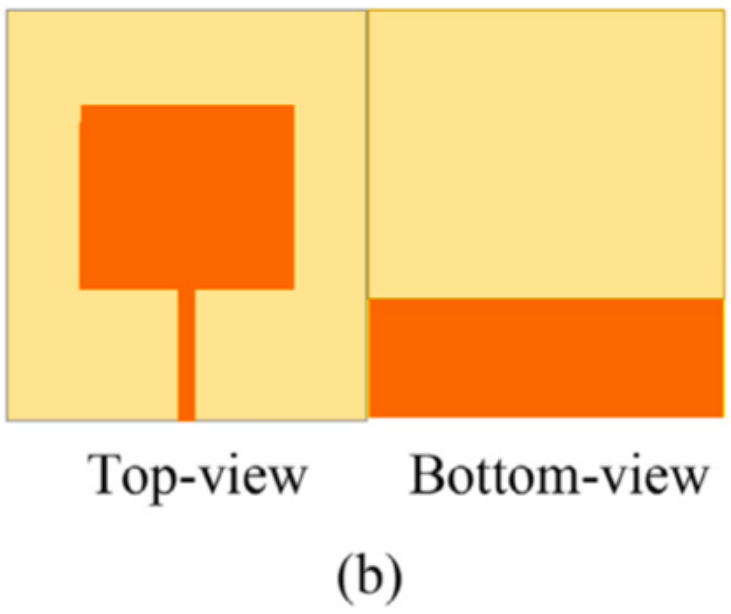

Figure 6:Stage 2. This is a partial ground as shown in the bottom-view. Shown in orange. The $S_{11}$ parameters is considerably reduced and produces an impedance bandwidth below -10dB which is obtained as a consequential ground height reduction.

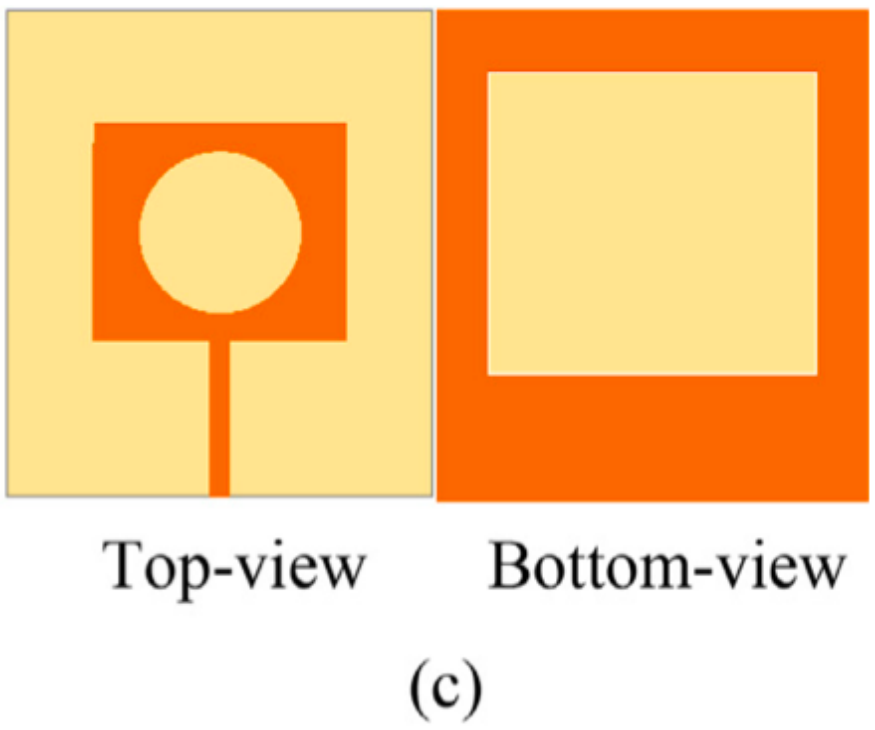

Figure 7:Phase 3. In this phase there is a circular patch but with slotted ground. The typical route of the surface current was interrupted with a circular patch and there was a rectangular shape at the ground plane which had inner slanted corners. $S_{11}$ below -10dB was achieved.

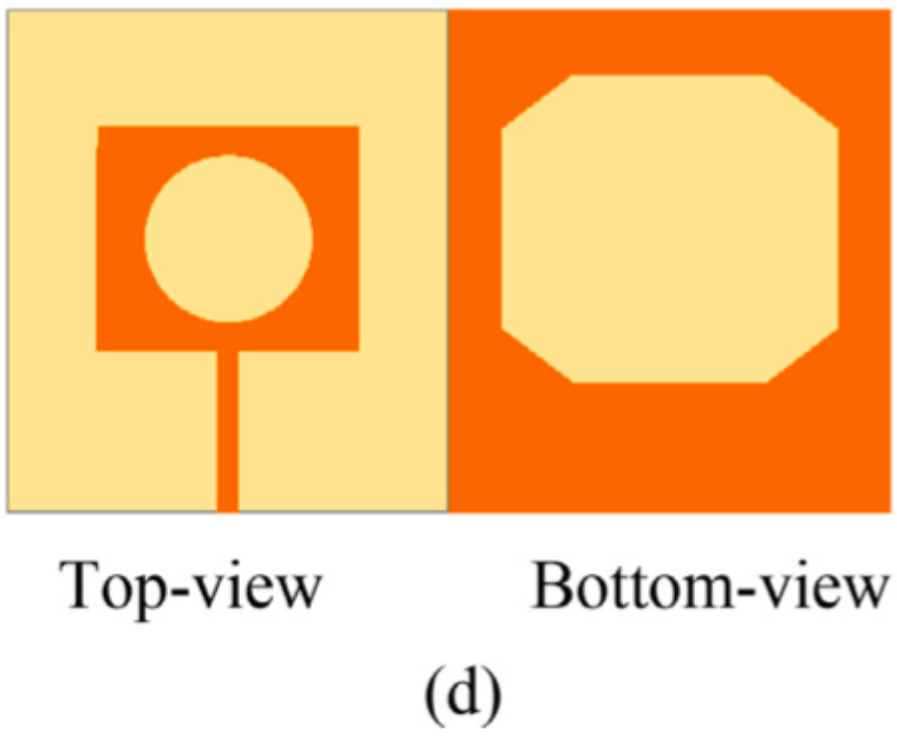

Figure 8:Phase 4. In this phase also there is a circular patch but with a different bottom shape structure. In this also the typical route of the surface current was interrupted with the circular patch centre. This ground shape and circular patch enables many resonance modes. The operating modes is from 1.2GHz to 2.7GHz. $S_{11}$ below -10dB was achieved.

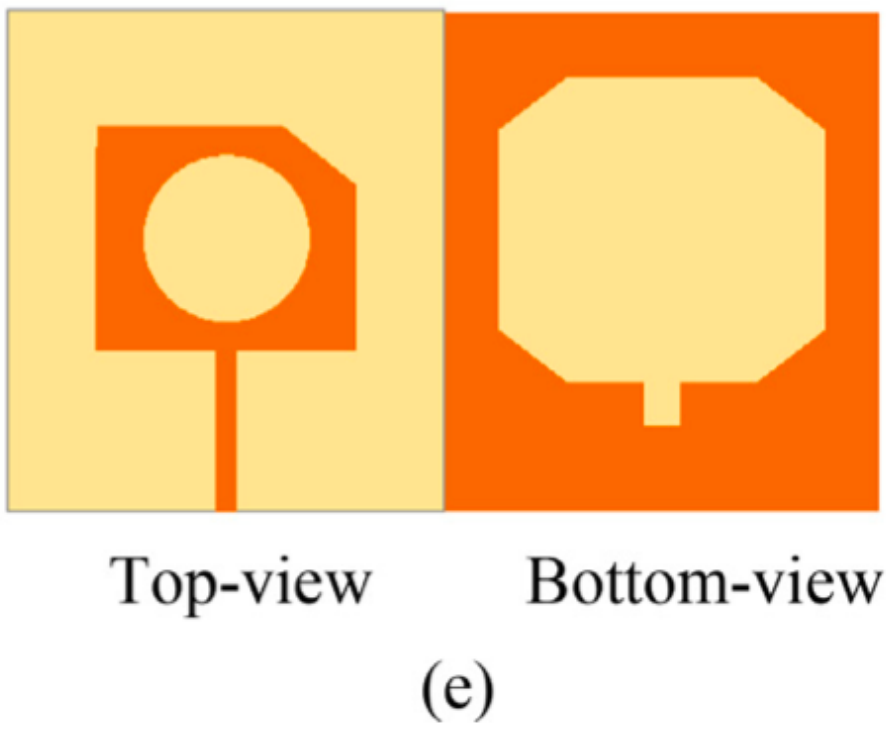

Figure 9:Phase 5.(proposed structure).There is a rectangular slot and a sloping hypotenuse shaped top one sided slot. This has resulted in changing the direction of the current in the setup. Pertinent spreading of the current in the ground plane. This results in the setup operating between 1.22GHz to 3.45GHz. The $S_{11}$ parameters were observed to be below -10dB.

Surface current mechanism. A reversed current is experienced in the ground plane as a consequence of the radiator in patch antenna. There is a slot introduced just under the radiating patch, this makes the surface current on the ground to loop around the slot. Making slots on the ground plane generates extra polarization, owing to surface currents.

The given below for surface current for various modification at 3.3GHz.

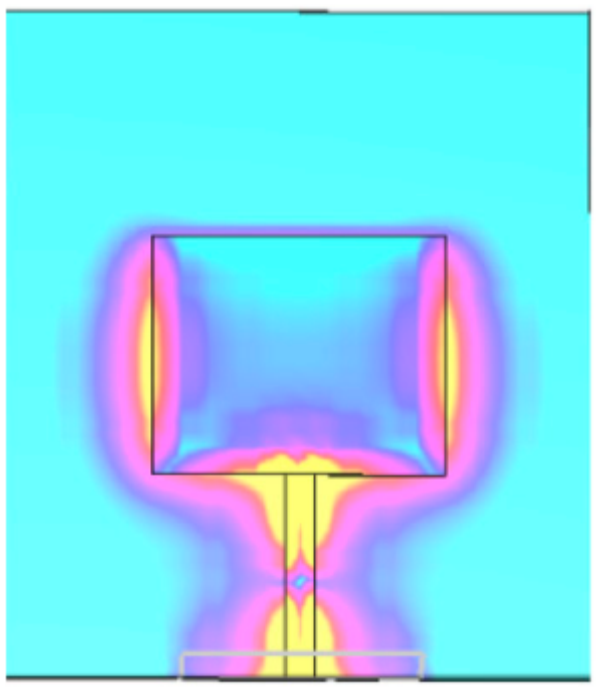

Figure 10:Stage 1.Current distribution. No effective bandwidth is achieved in phase 1 as significant amount of current is circulated around the edges and feedline of the radiating patch. The unslotted element could not expand the current flow.

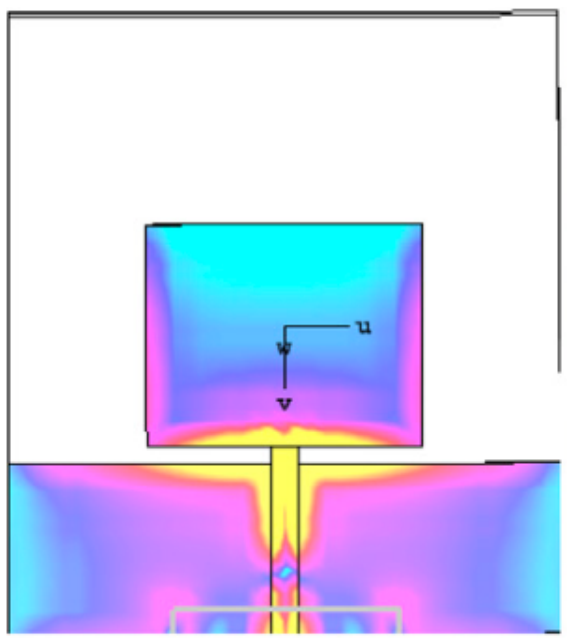

Figure 11:Stage 2. Current distribution profile. The fragmented ground plane enhanced the current flow on the ground plane. There were generated field radiated effectively from the ground plane. This design aided in the $S_{11}$ to be below -10dB.

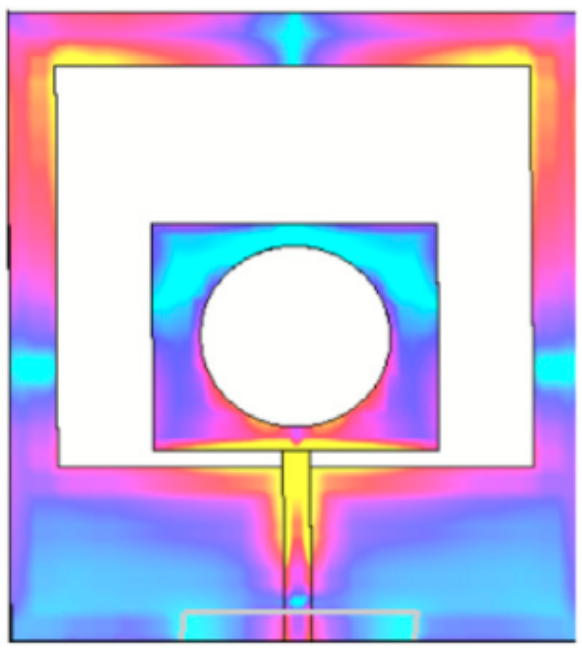

Figure 12:Stage 3. In this stage the circular slot patch and the rectangular slot on the ground enhanced the dissemination of the surface current on the ground by reducing the form factor which is the amount of energy is getting passed through the surface from one end to another. This resulted in multiple resonance modes.

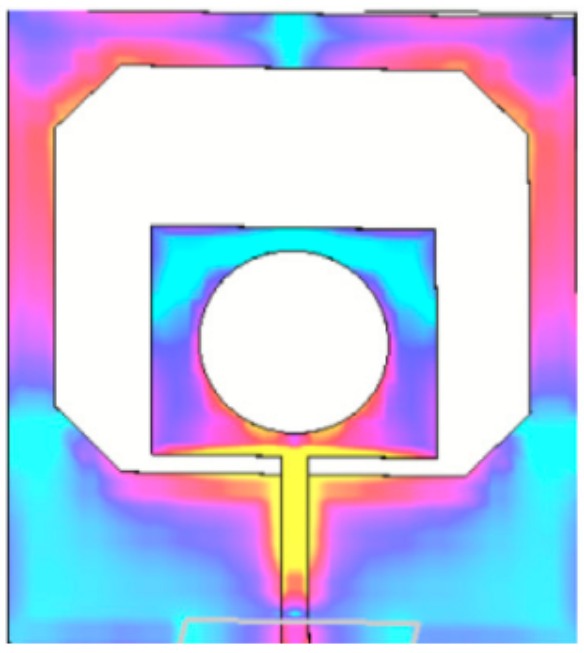

Figure 13:Stage 4. The current distribution was altered by the rounded base. This allowed broad operational band between 1.25GHz to 3GHz range.

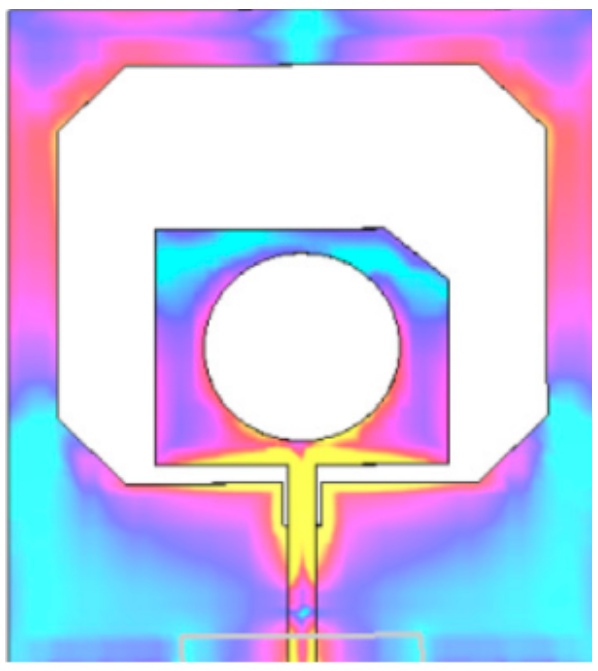

Figure 14:Stage 5. In the last phase another rectangular slot in the ground place was introduced this resulted in the increase in the ground plane useable region, this resulted in making the route of the current flow longer than other cases. Further enhancing the bandwidth and frequency.

Parametric optimization was performed of the antenna design. The values of the various parameters were adjusted using optimization. Analysis of gap between radiating patch and ground G is considered. This is the gap between the radiating antenna and the brain head phantom. This is done to obtain optimal impedance matching. This is done to achieve proper antenna performance. G was varied from 0.5 to 2.5mm, the best obtained results was for G=1.5mm.

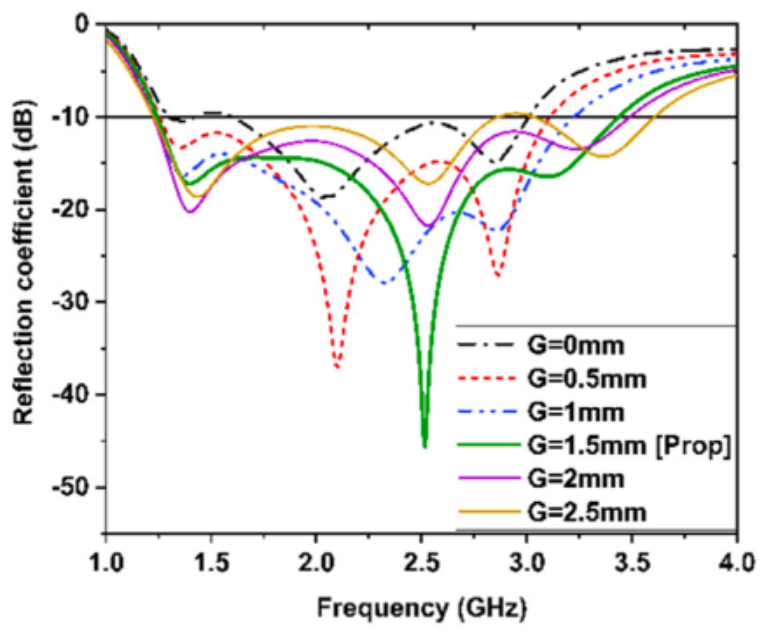

Figure 15:Reflection coefficient for different values of Gap G. The proposed design with G=1.5mm shows optimal performance.

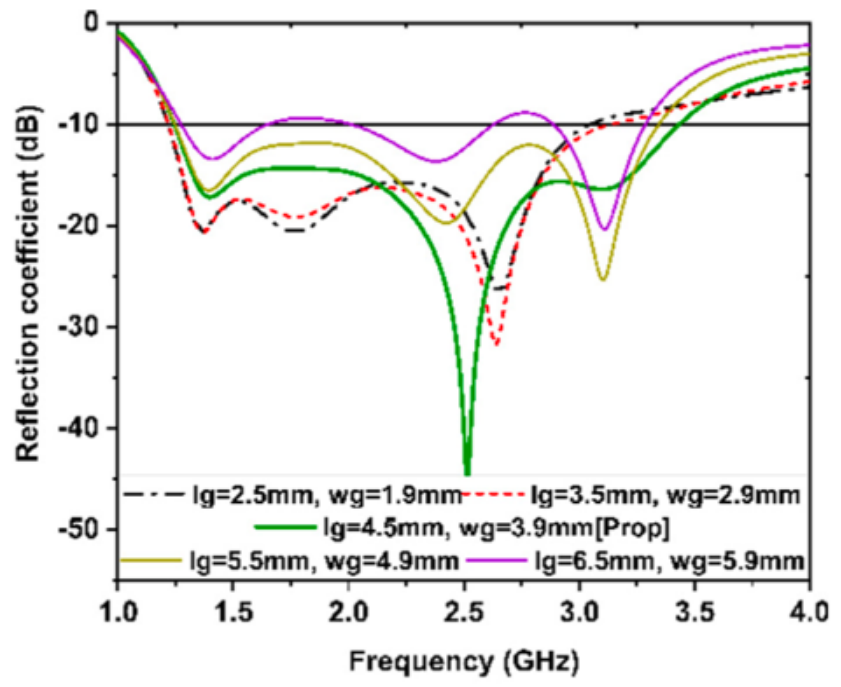

Figure 16:The values of $W_g$ and $L_g$ was varied to check the bandwidth. Optimal performance bandwidth of 2.23GHz was achieved for values of $L_g$=4.5mm and $W_g$=3.9mm. For other values of $W_g$ and $L_g$ the reflection coefficient and bandwidth was tested, the values of bandwidth obtained was affecting the performance of the antenna, by the fact that for values of $L_g$=3.5mm and $W_g$=2.9mm the reflection coefficient did not show a fall more than -30dB as compared to the earlier taken values, any value close to -10dB affects performance.

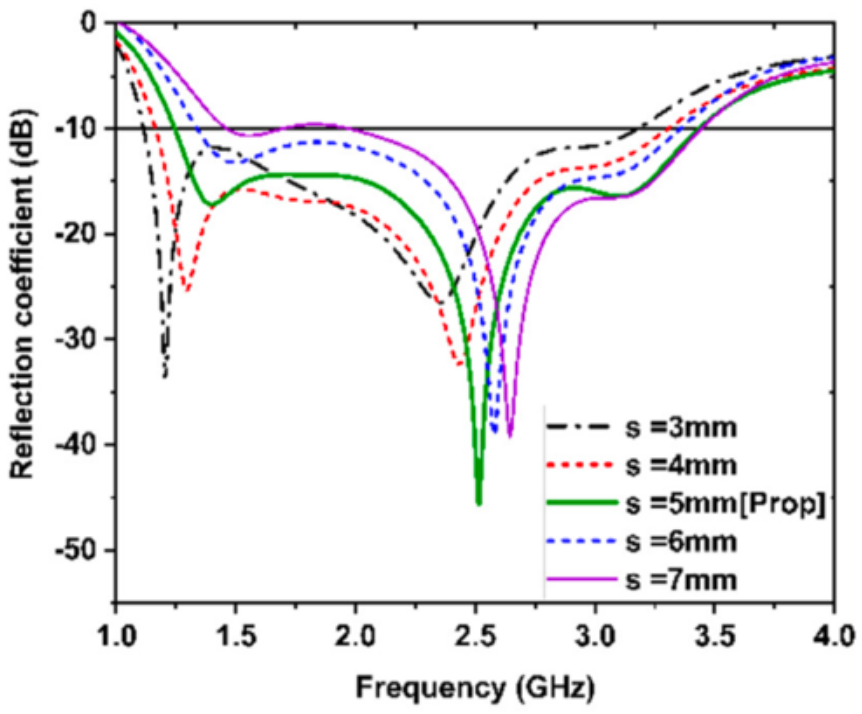

Figure 17:The values of lateral edge width was calculated S was varied between 3mm, 4mm, 5mm, 6mm, and 7mm.

S=5mm the obtained values was calculated to be optimum.(proposed design).

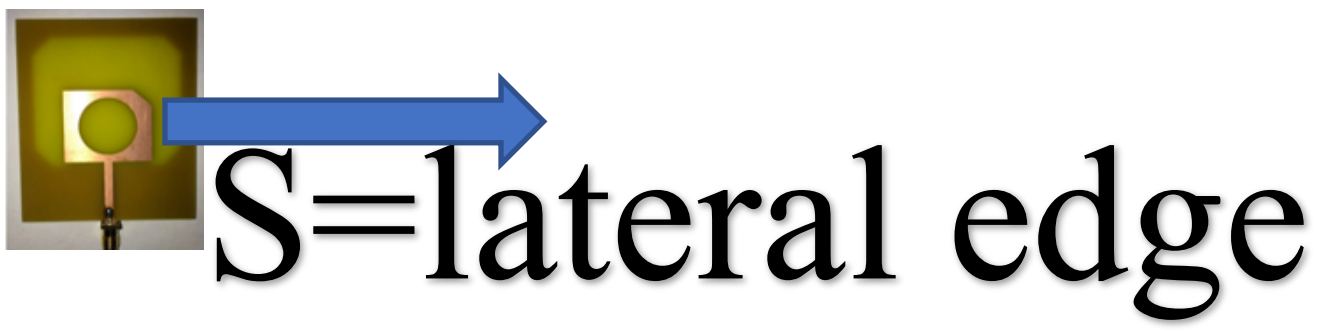

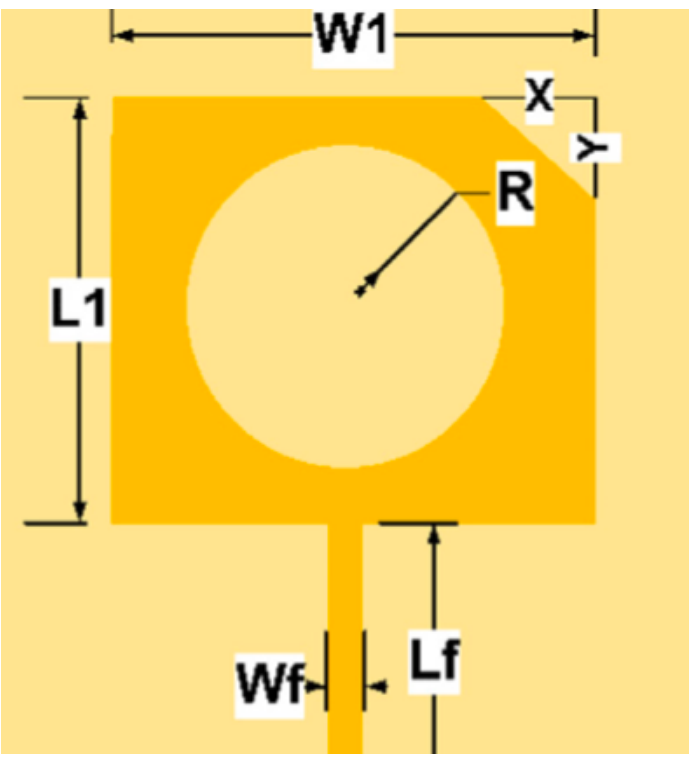

Figure 18:The circular slot radius R. As in the design there is a square from which there is a circular slot cut out. The radius of the circular radius R is varied.

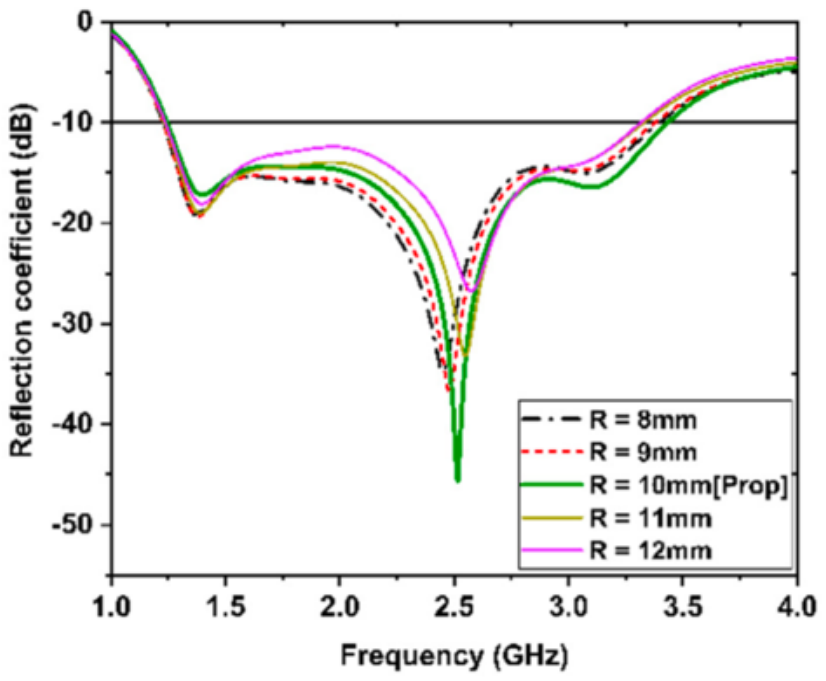

Figure 19:The radius R was varied between 8mm to 12mm with increments of 1mm. Showed best values for R=10mm(proposed)[Green line]

-45dB reflection coefficient made the proposed design good.

The measured antenna has an efficiency of more than 85%. The antenna was tested for various applications as broadband antenna, etc. The realistic head Hugo model was designed. The mode consisted of six different head tissues, which were fat, skin, cortical bone, cerebrospinal fluid, white matter and grey matter. When the electromagnetic waves was sent into the skull for imaging, a rise in EM field absorbance was absorbed into the human organic tissues.

In the second paper Microwave brain imaging system to detect brain tumour using metamaterial loaded stacked antenna array by Amran Hossain et.al they have used a similar array configuration as the first paper but they have used a metamaterial in its design. They have used a three layered meta material stacked structure. They have used low priced rogers RT5880 and RO 4350B substrate, they have used optimization to find the best dimensions to be 50X40X8.66mm$^3$.

Design steps:

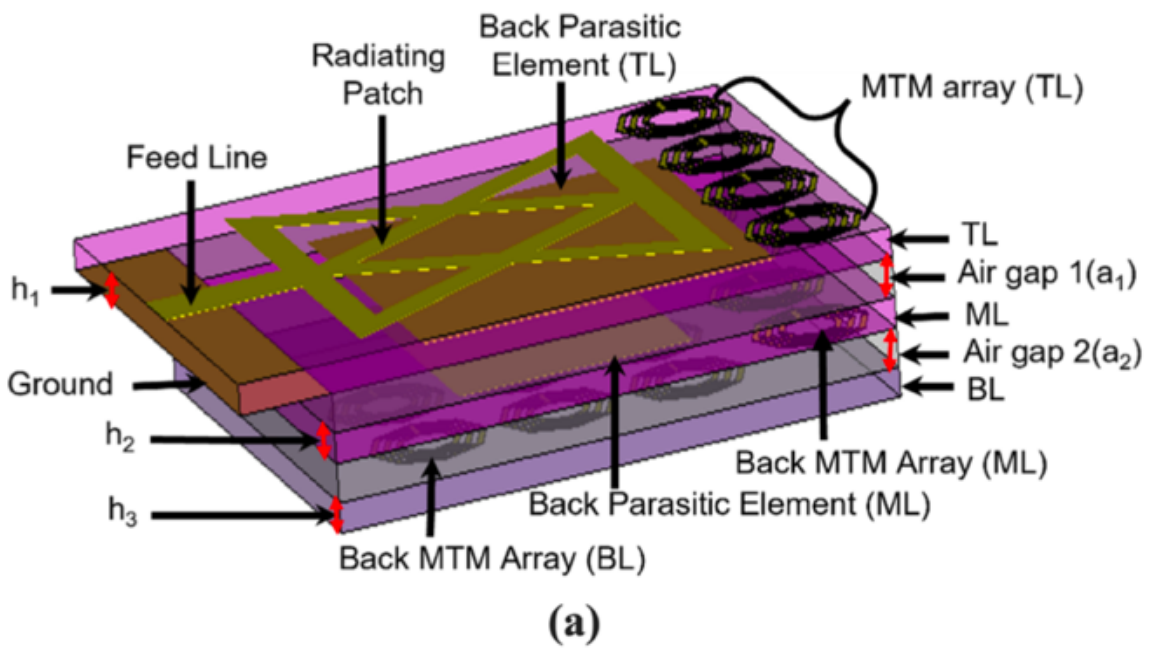

Figure 20:Metamaterial loaded 3D structure is shown as follows. Top layer, middle and bottom layer is shown. A 2mm air gap is considered between each of the bottom, middle and top layer. The main radiating patch and feed line is designed on the top layer. A 50ohm impedance matching is done to the feedline connecting the radiating patch.

The design is of spider net shaped array on the TL,ML and BL to enhance the antennas performance.

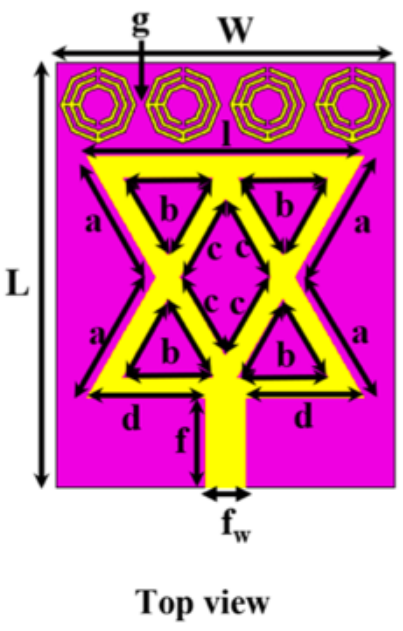

Figure 21:This is the top layer top view. They have used four spider shaped radiating patch designs. This is a 1X4 MTM array.

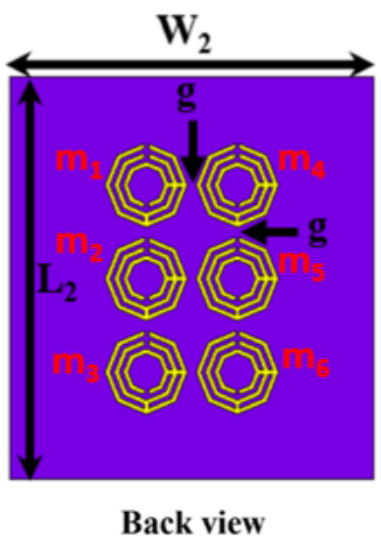

Figure 22:3X2 MTM array is used on the bottom layer to enhance the gain and directivity of the antenna. When the operating frequency is below 1GHz the obtained image is unclear and fuzzy.

A major radiating patch and feed line is designed on the top layer.

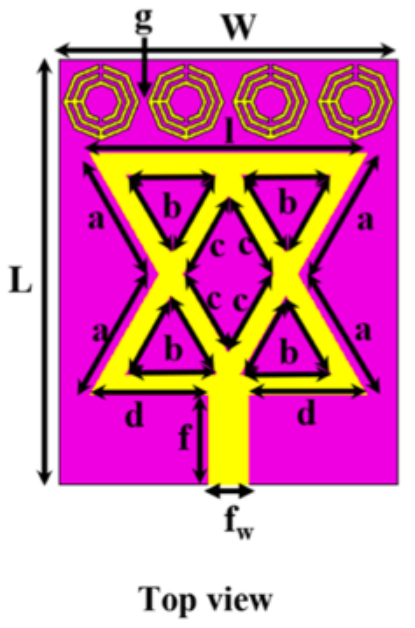

Figure 23:This is the top layer. Two triangles are designed opposite to each other as shown in the figure as radiating patches. There is a feed line in the top layer.

This is a wideband antenna with wider or large band of frequencies, in narrow band antenna the range of frequencies are narrow.

Design geometry specifications: It so happenes that when the antenna operates in this wideband the electromagnetic waves are made to easily penetrate through the brain tissues and the reflected waves are made to collect to the receiving antennas. In the design there is TL, BL, and ML with air gaps which are 2mm between. In the TL there is a radiating patch as two triangles inverted with each other.

$$L_p = \frac{2c}{3fr\sqrt{\mu reff}} \qquad (1)$$

$$f_w = \frac{7.48\ X\ h}{e^{(Zo(\frac{\sqrt{er+1.41}}{87})}} - 1.25\ X\ t \qquad (2)$$

$$\mu_{reff} = \frac{\mu r+1}{2} + \frac{\mu r-1}{2} + [1+12\frac{h}{W}]^{-1/2} \qquad (3)$$

$$F_l = \frac{c}{2Xl\sqrt{e\ eff}} \qquad (4)$$

Patch width($L_p$), feed line width($f_w$), lowest operating frequency($f_l$) are calculated.

The $S_{11}$ parameters reflection coefficient.

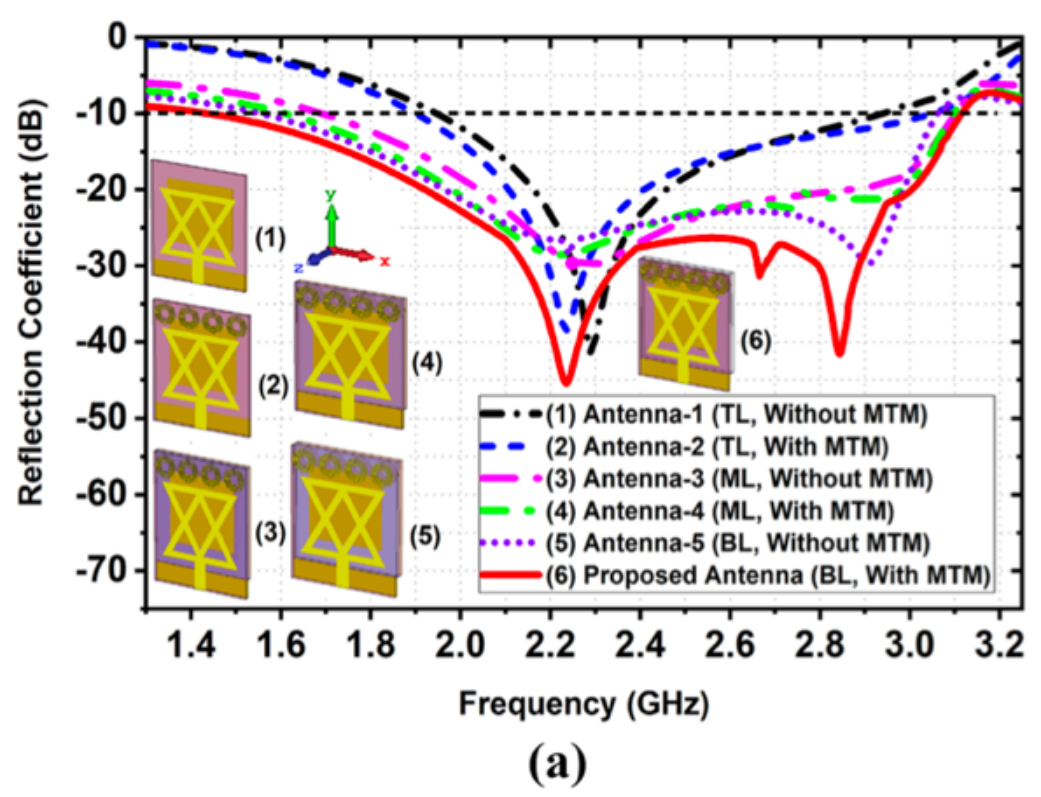

Figure 24:The proposed design with MTM layer shows the $S_{11}$ reflection coefficient way below -10dB and touching -50dB in the red line as compared to the other designed which had antenna-5, purple line, BL without MTM, antenna-4 , green line, ML with MTM, antenna-3, pink line, ML without MTM, antenna-2 ,blue line ,TL with MTM, and antenna-1 , black line, without MTM) .

There are BL,ML and TL and air gap between each ML and BL are better performance achieved by the overall structure.

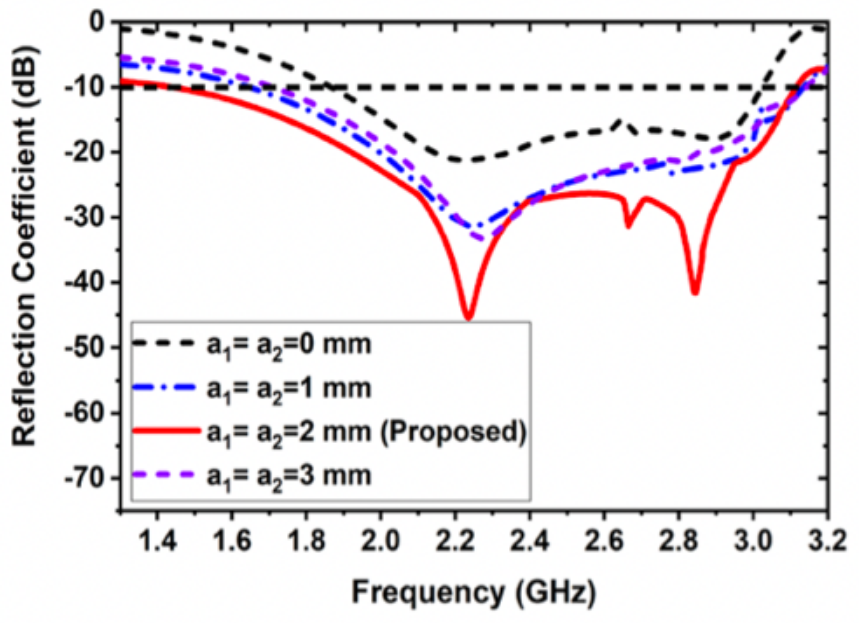

(a)

Figure 25:Reflection coefficient values of a for no air gap spacing i.e for a=0mm(black line), for a=1mm(blue line), proposed a=2mm(red line) and a=3mm for purple coloured line.

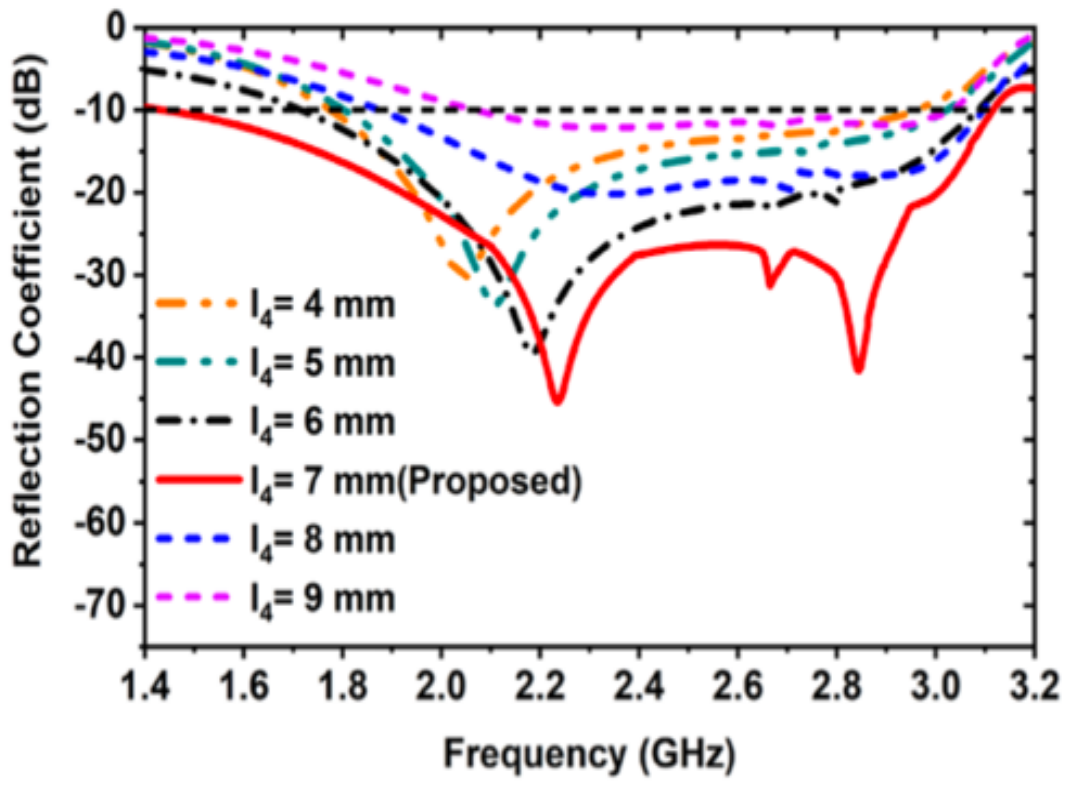

(b)

Figure 26:The length of the ground plane is varied keeping all the other elements constant. For $l_4$=4mm the the obtained operating band is 1.77-2.85GHz and a resonance of 2.05GHz. in the second case when $l_5$=5mm the working band is 1.71-3.08 GHz and a sonority at 2.18GHz, in the third case when $l_6$=6mm, the attained operating band is 1.71-3.08GHz with a resonance at 2.18GHz, In the fourth case when $l_7$=7mm the achieved frequency is better than all the other measured lengths.

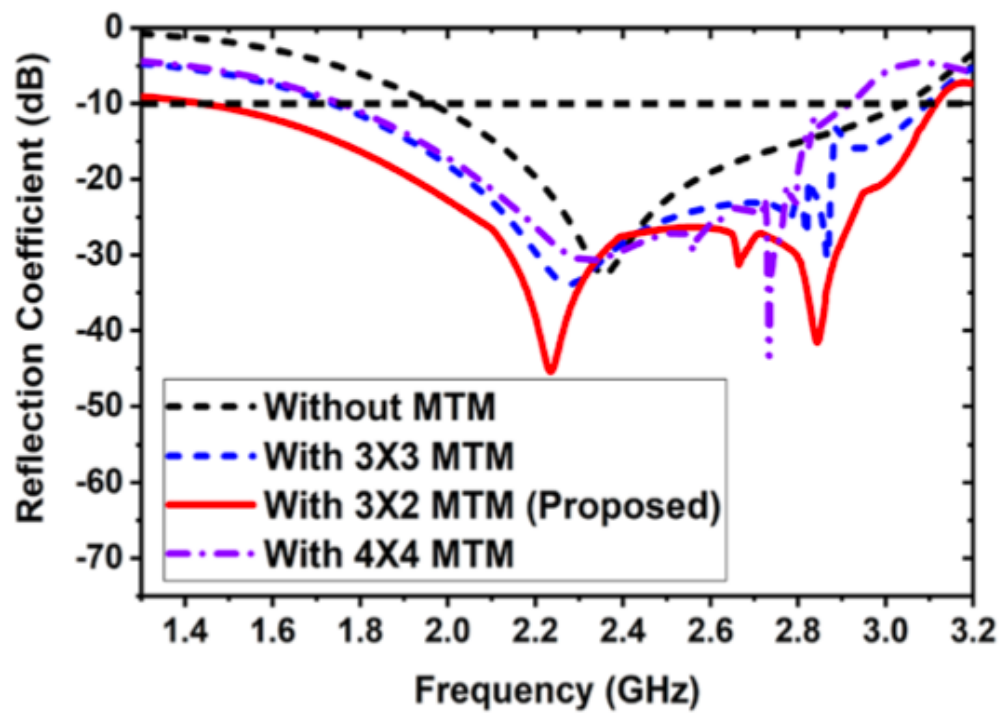

(c)

Figure 27:When 3 X 3 MTM is used , shown in blue colour, the antennas frequency of operation is 1.73-3.09GHz with two resonances at 2.27 and 2.86GHz, making the reflection coefficient very high. Compared to 4X4 metamaterial, 3X2 metamaterial shows good results with resonance frequencies at 2.24GHz, 2.66GHz and 2.84GHz respectively.

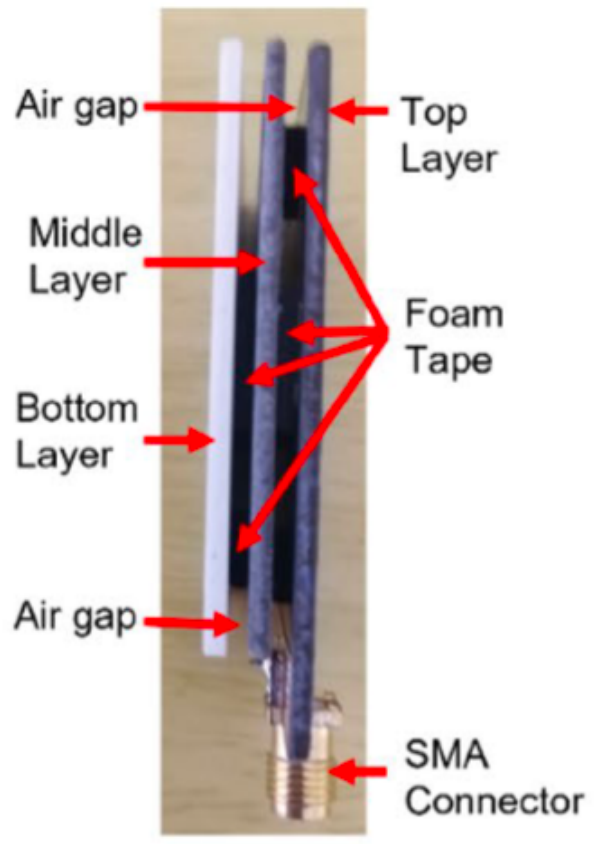

Figure 28:In this antenna, there is a Top layer, bottom layer and a middle layer. The TL and ML and BL are connected with each other with a 2mm foam tape which is double sided. Feed is connected to the top layer.

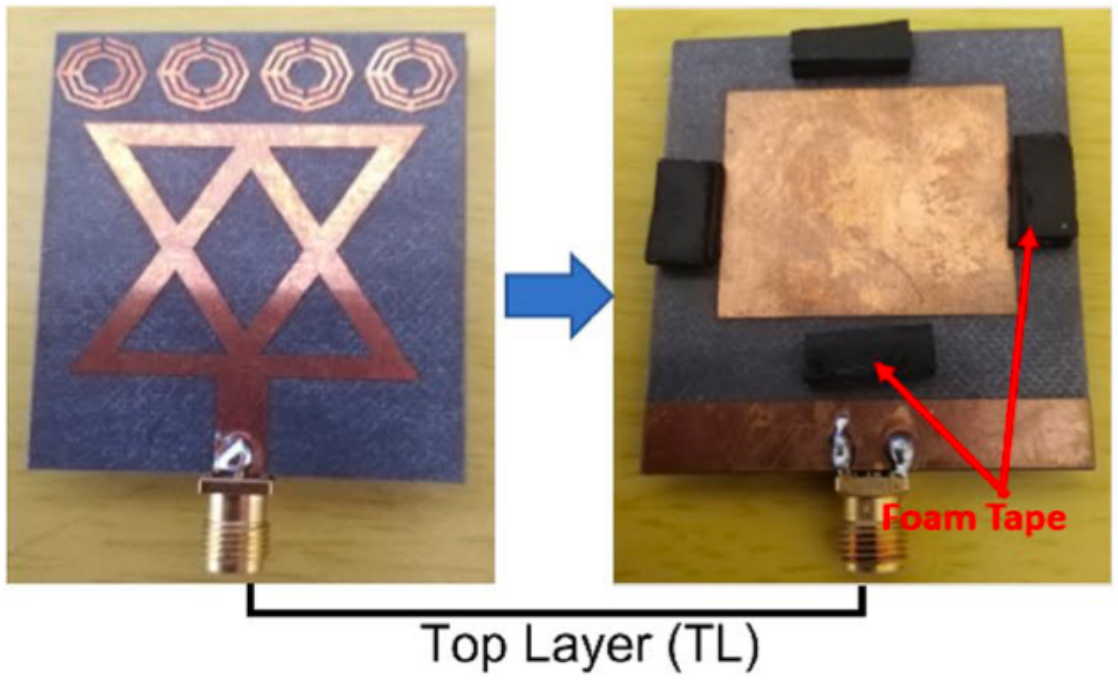

Figure 29:The top layer is fabricated as shown above. There is a feed in the top layer. This is a plate top later with 2mm foam at the back of the layer.

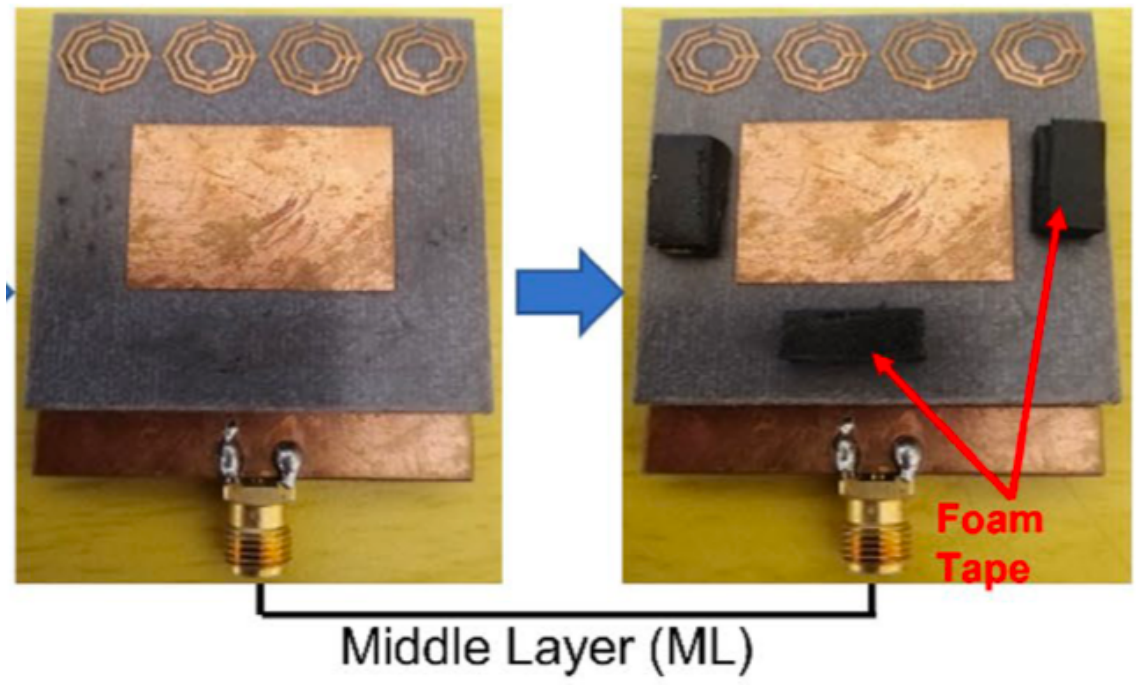

Figure 30:The middle layer is shown as follows, connecting the bottom layer and top layer with a foam tape.

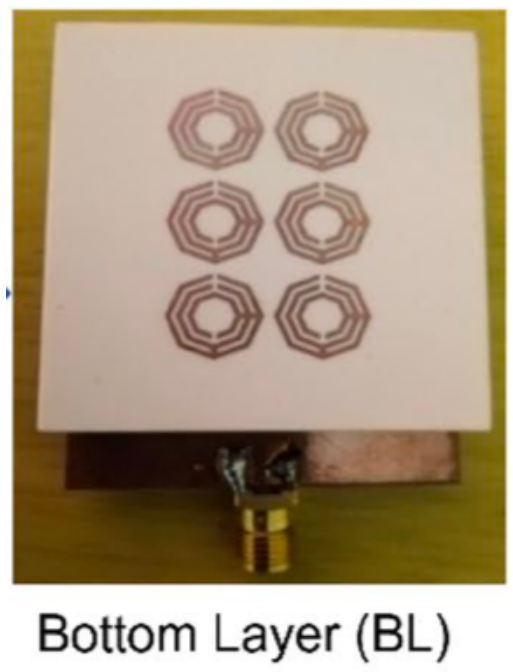

Figure 31:The bottom layer is shown as follows, there is a spider shaped array of antenna structure printed on the bottom layer. This is the back side of the completed antenna.

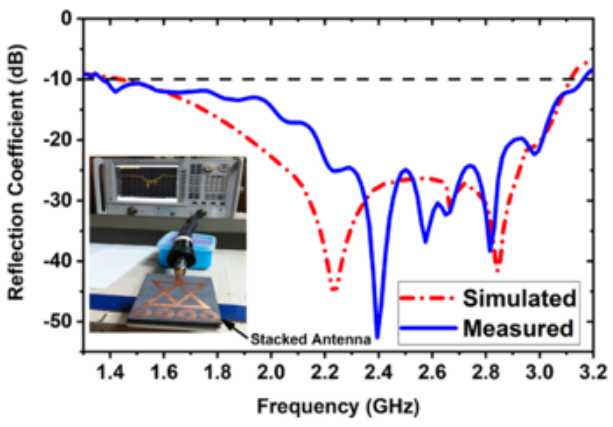

(a)

Figure 32:Reflection coefficient and frequency curve for measured and simulated results. The measured results show greater $S_{11}$ parameters.

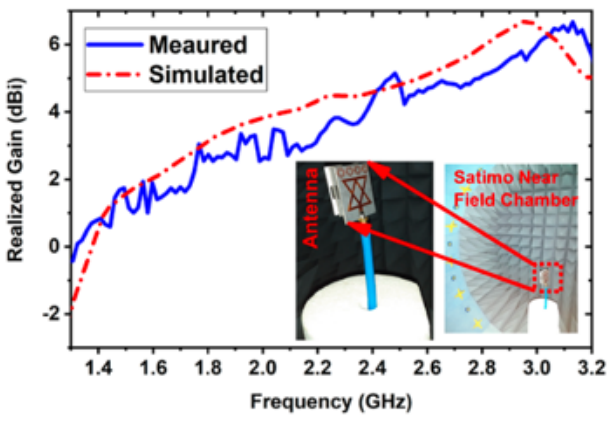

Figure 33:The realized gain versus frequency graph is obtained as follows. There is a coherence obtained between the measured and simulated results.

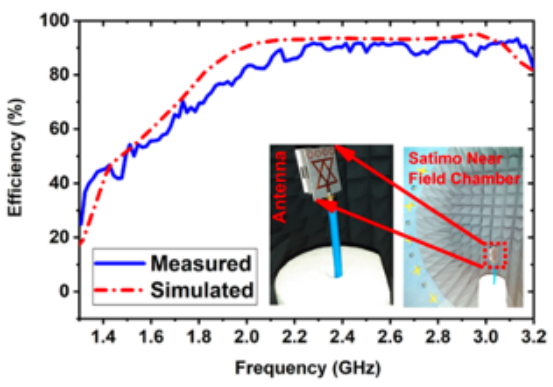

Figure 34:The efficiency of the antenna versus the frequency is plotted as follows. There is a coherence obtained between the measured and the simulated results.

Many antennas have been developed in the range of frequency 1-4GHz but in the totality they have failed to develop a clear, noise and fuzzy free image from the $S_{11}$ parameters. The proposed MTM stacked array antenna achieved high gain and efficiency and directional radiation properties as compared to similar antennas such as conformal and foldable antennas. The antennas also achieved high fidelity factor in which the waves were not distorted during receiving of about 98%. The signals could penetrate through the head tissue with great efficiency. There was drastic increase in the radiation characteristics of the antenna due to the properties used in the design of the antenna. It is observed that the back lobe is reduced significantly due to the usage of the MTM on the stacked layers of the antenna. Moreover in the radiation characteristics of the antenna when observed between 1-4GHz, they were more directive at higher frequencies as compared to the lower frequencies.

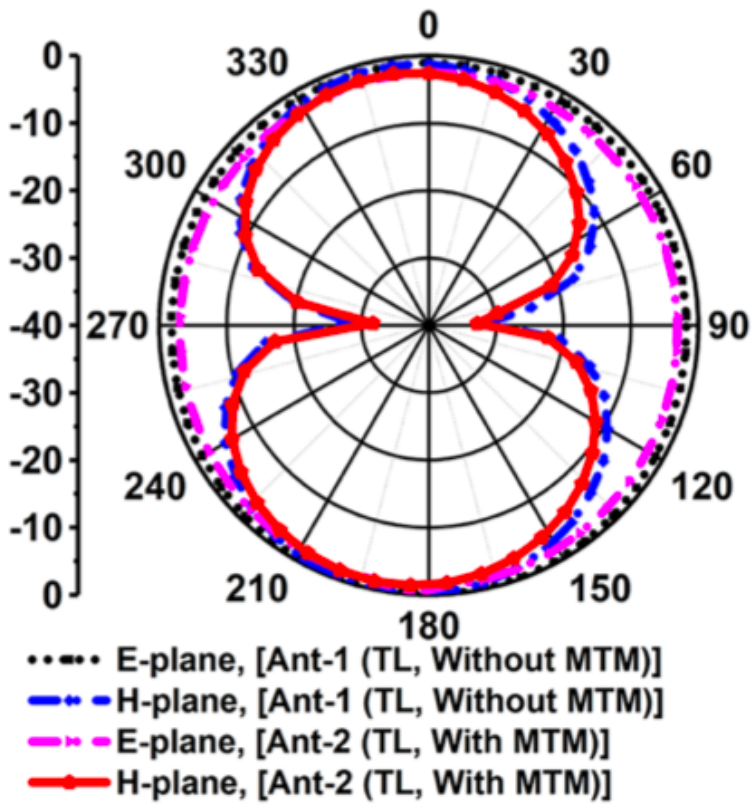

Figure 35:The radiation pattern obtained in the H plane shows concurrently more beamwidth in the boresight direction as compared to the E plane. This signifies that the antenna mostly radiated in the -Z direction.

The simulated and measured far field radiation pattern is shown as follows in the figure below.

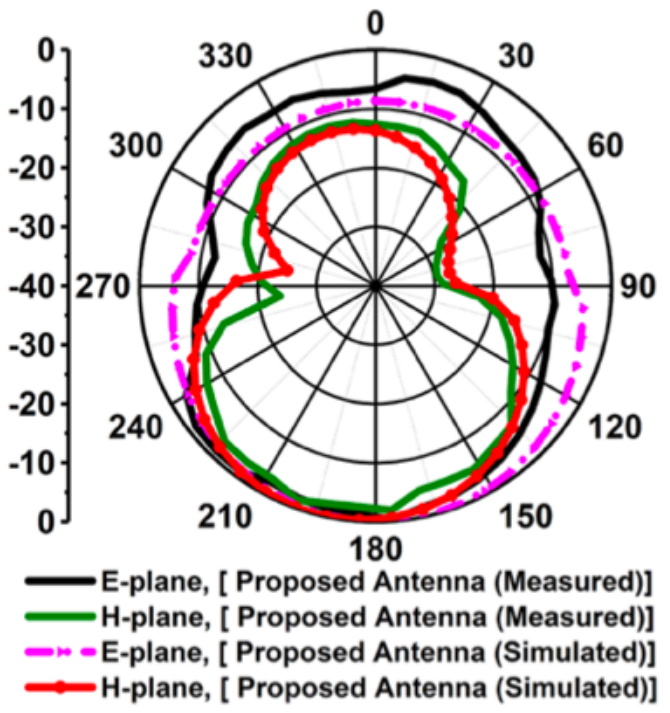

Figure 36:Results showed Good agreement between the measured and simulated far field pattern. The antenna will be placed at some distance from the antenna(very small) hence it will require the near field characteristics of the antenna.

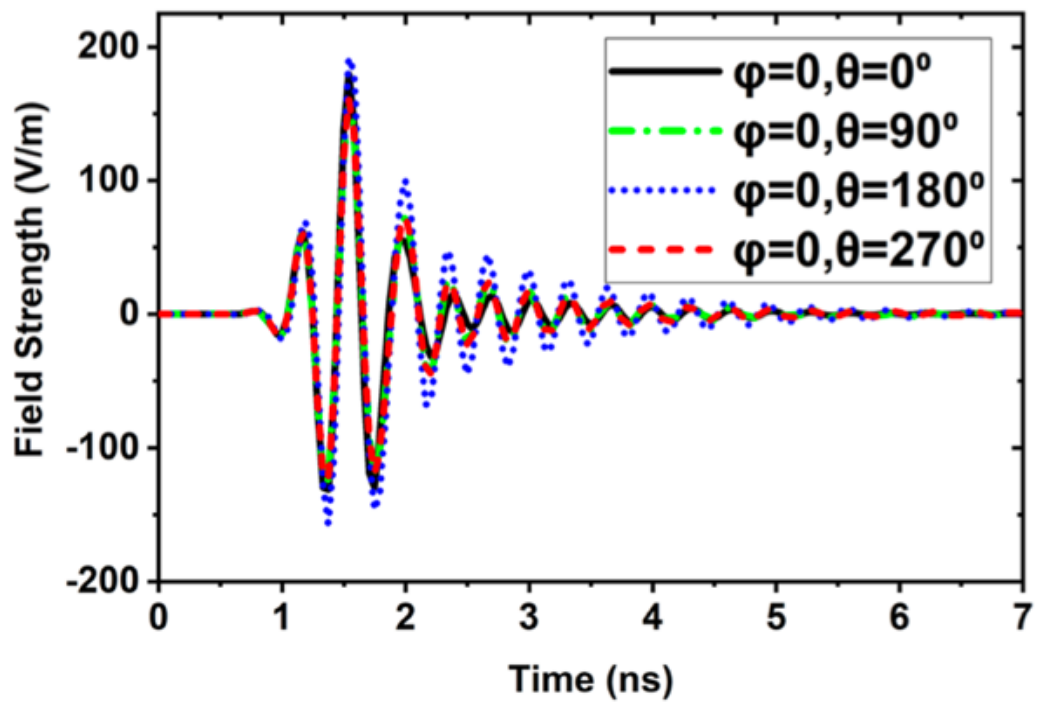

Figure 37:The near field radiation pattern of the antenna is given as follows. The antenna will be kept at close distance from the head phantom model hence we try to find the near field characteristics of the antenna.

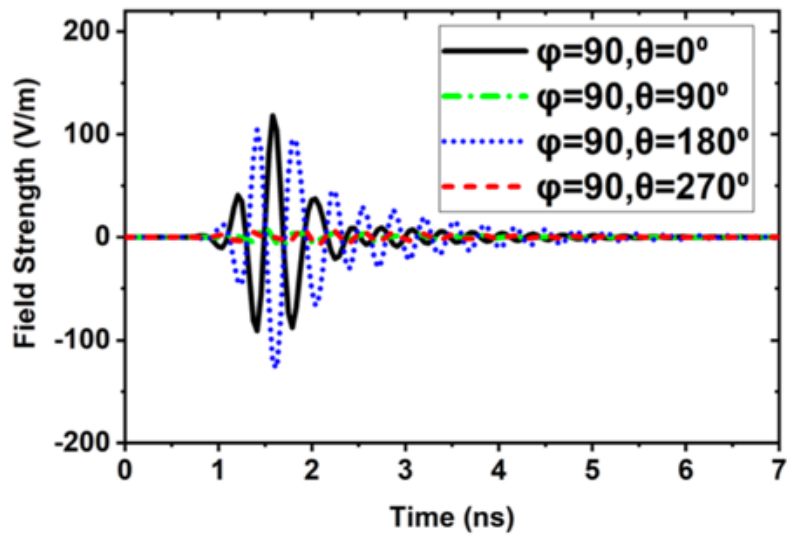

Figure 38:Near field pattern in the H plane is observed to show the radiation pattern as follows.

The Fidelity factor is used to calculate the near field performance by using the following methods.

(1)Using the E field probs

(2)By using the transmitting and receiving antennas by considering the near field distance between in few scenarios. In such cases the fidelity factor(FF) is used to investigate the near field characteristics, i.e. the distortion in it. This is defined by the correlation coefficient between the received radiated fields pulses, $E_{rad}$ in various directions and the gaussian plane which is excited at the input of the antenna. Specially in this work, the finite difference time domain(FDTD) method is used to find out the degree of distortion in each of the defined methods. The proposed antenna performed better in far field and in near field. The scattering parameters are observed for the antenna array model with the imported head phantom model.

In this paper research they have fabricated a six layered head phantom model to investigate the antenna performance. The output S parameters are seen for benign and malignant tumours. There is a nine antenna array setup in which one antenna acts as the radiating antenna and the rest eight antenna act as the absorbing antennas.

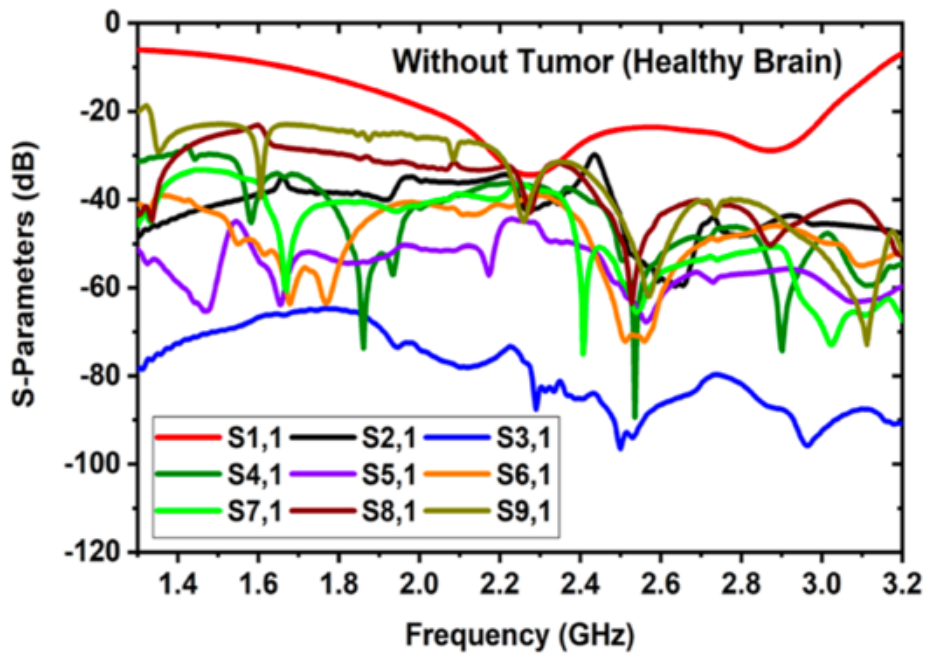

Figure 39:The Scattering parameters without tumour, that is for a healthy brain is given as follows. There is no distortion of the back scattered signals in case of the healthy brain observance of scattering parameters.

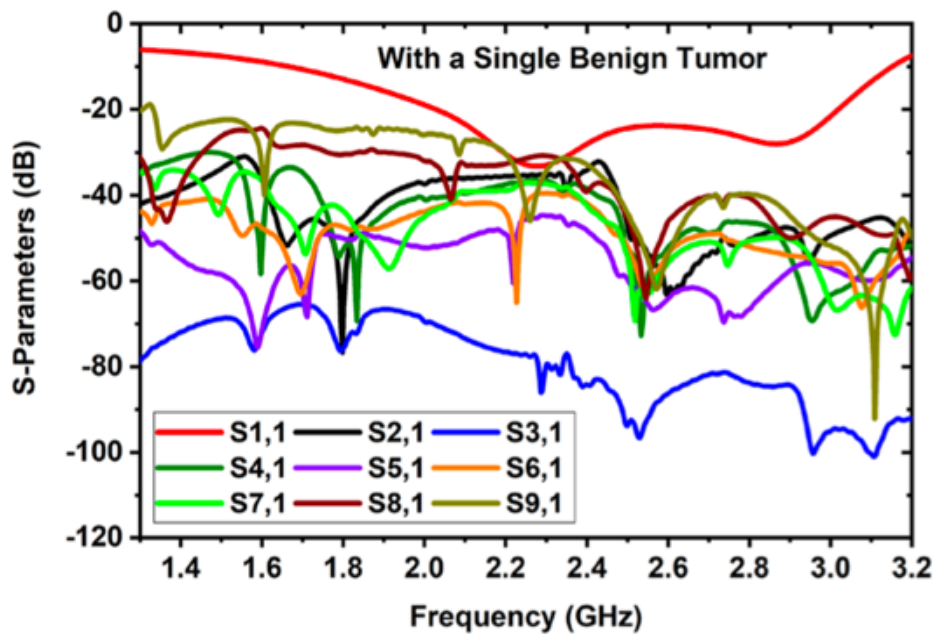

Figure 40:With a single tumour there is distortion of signals to some extend as compared to the healthy brain.

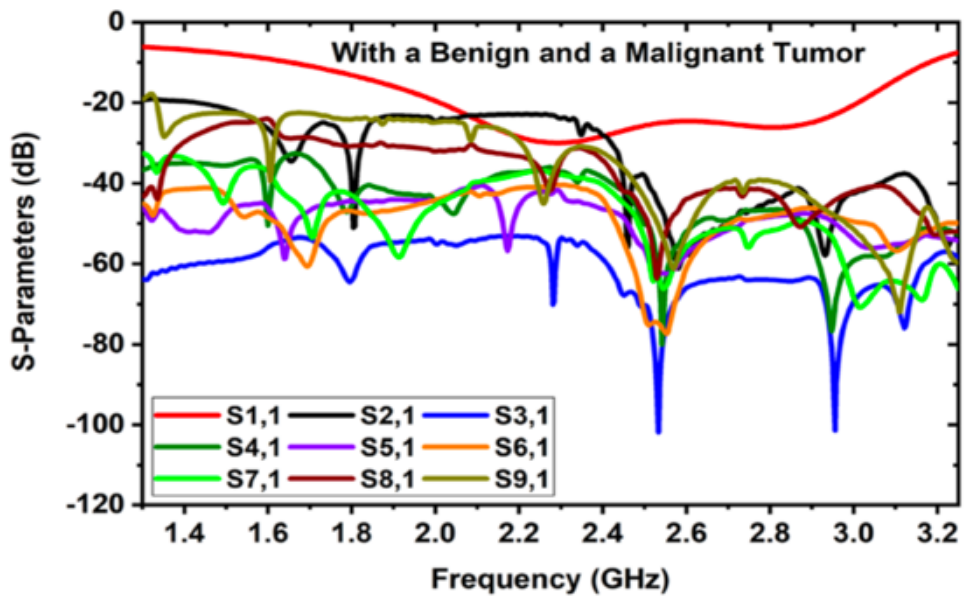

Figure 41:For a malignant and benign tumour(single) the distortion is shown as follows. There is maximum distortion in case of this observance.

It is observed that permittivity is decreased with increase in frequency while conductivity is increased with increase in frequency.

Discussion about the microwave imaging results:

A head phantom with six materials is fabricated(together with benign and malignant tumours). Benign tumour is in circular shape while the malignant tumour is in the elliptical shape. Permittivity decreases with increase in frequency, while the permittivity is increased for the same frequency(1-4GHz).

The proposed system the apparatus consists of MTM stacked nine antenna array, a custom shaped half-cut elliptical shaped helmet, a stepper motor, a portable stand, RF switch, microcontroller, and a PNA E8358 transceiver. In this setup the stepper motor is attached to the portable stand, which is rotating in the clockwise direction with 7.2 degrees angle at every step which is required the complete 360 degree dimensions of the head phantom. The helmet is applied with the motor with the motor shaft. The diameter of the helmet is 250mm. The antenna is attached inside the helmet with the help of double sided foam tape. The angular distance separation between each antenna is 40 degrees. Phantom model is placed at the centre of the helmet. The signals are further filtered to obtain the image by iteratively corrected coherence factor delay multiply and sum(IC-CF DMAS).

In the third paper we have reviewed Flexible Meander-Line Antenna Array for Wearable Electromagnetic Head Imaging by Alqadami et.al. One of the simple designs of antenna is the meander-line antenna.

Figure 42 and 43:

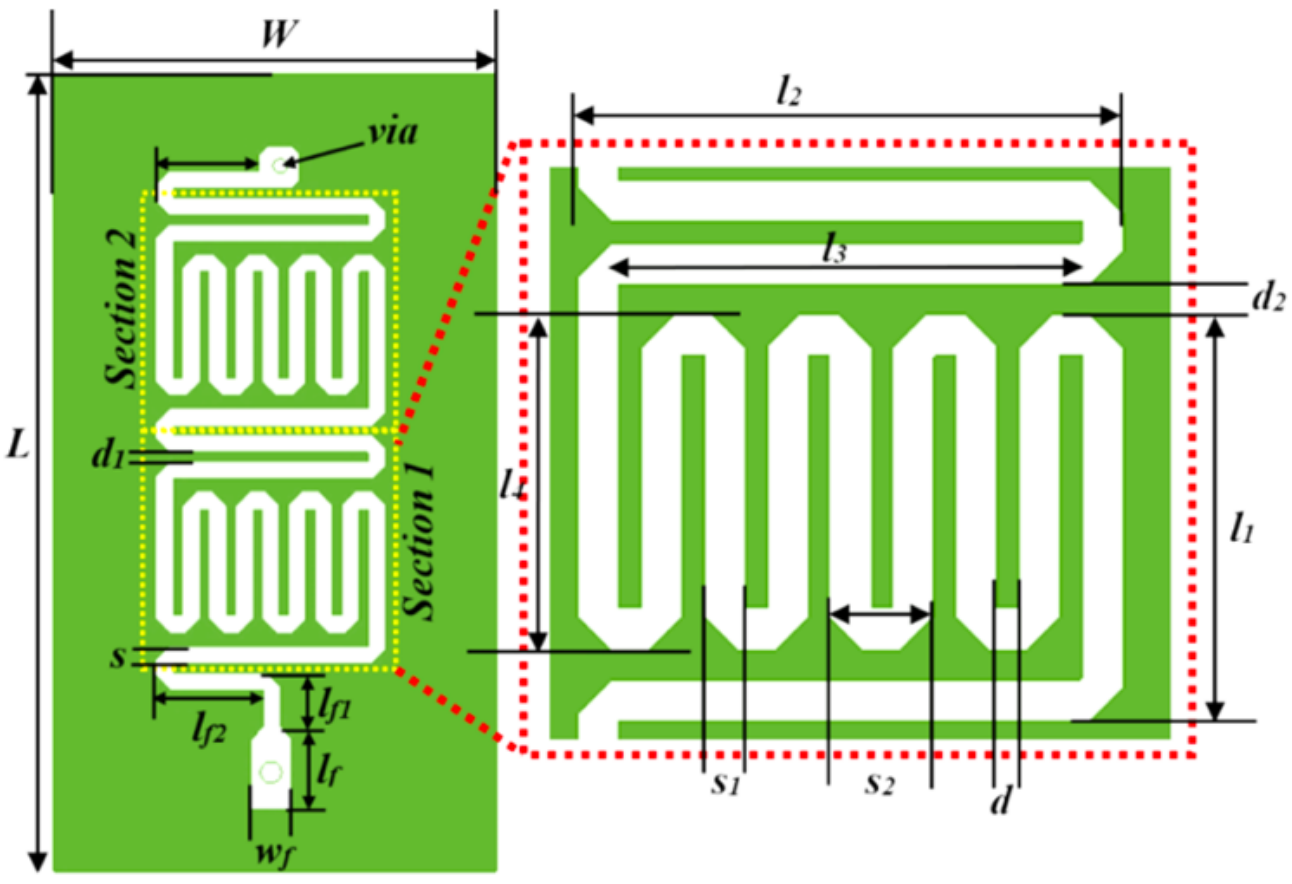

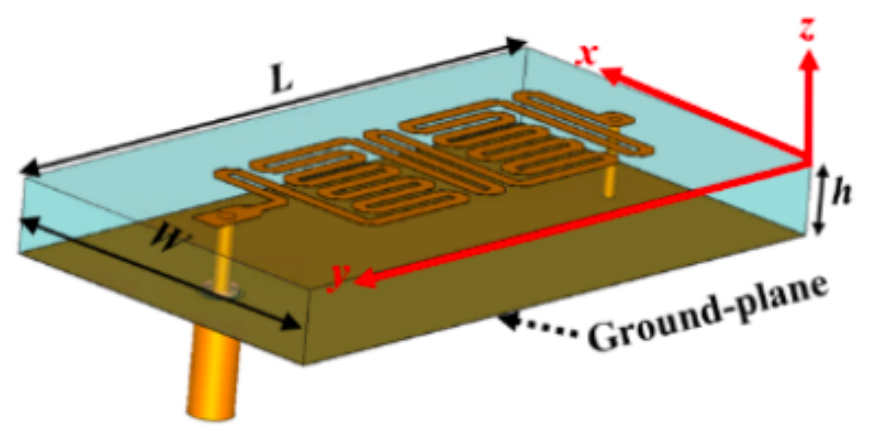

(a)

The antenna is designed on a RTV silicone rubber substrate with a thickness of 4mm. The antenna consists of a meander line, a 0.025mm thickness ground copper plane at the rear end, and a 50 ohm coaxial feed line. There are two symmetrical sections in the meander line. Section 1 and section 2. Each section contains nine vertical and three horizontal segments.

Figure 44 and 45:

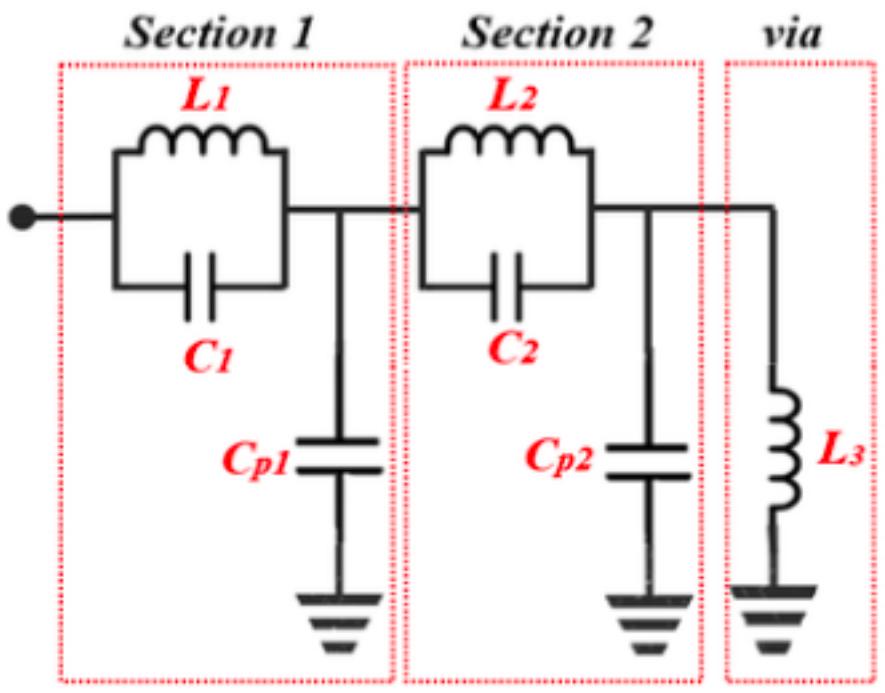

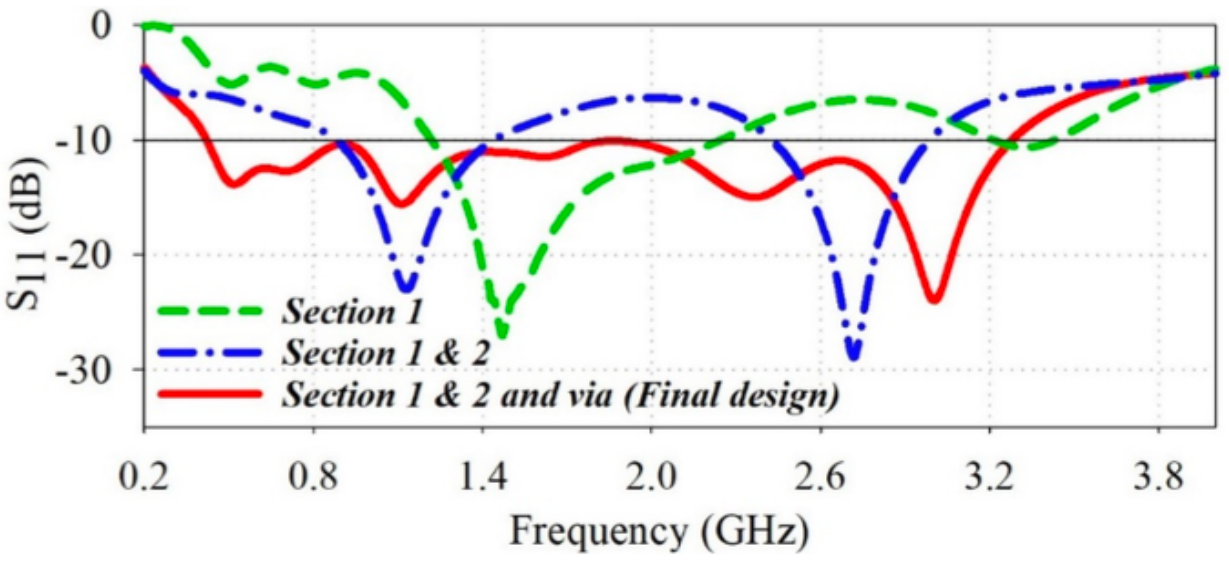

Figure 46 and 47:

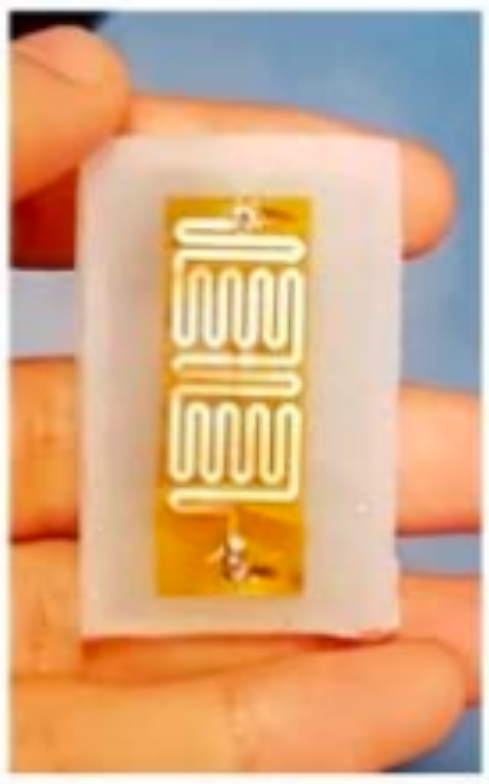

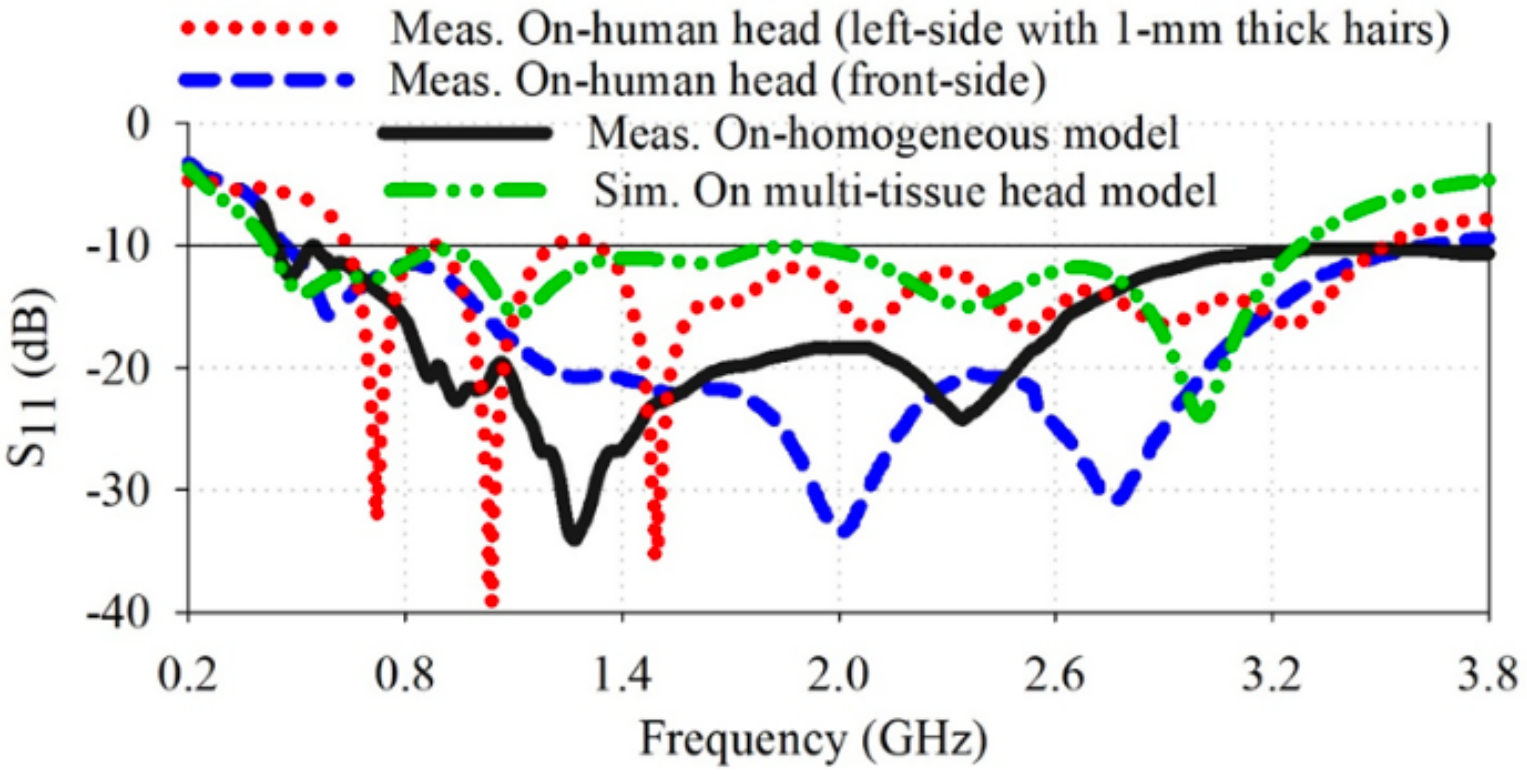

Figure 48:

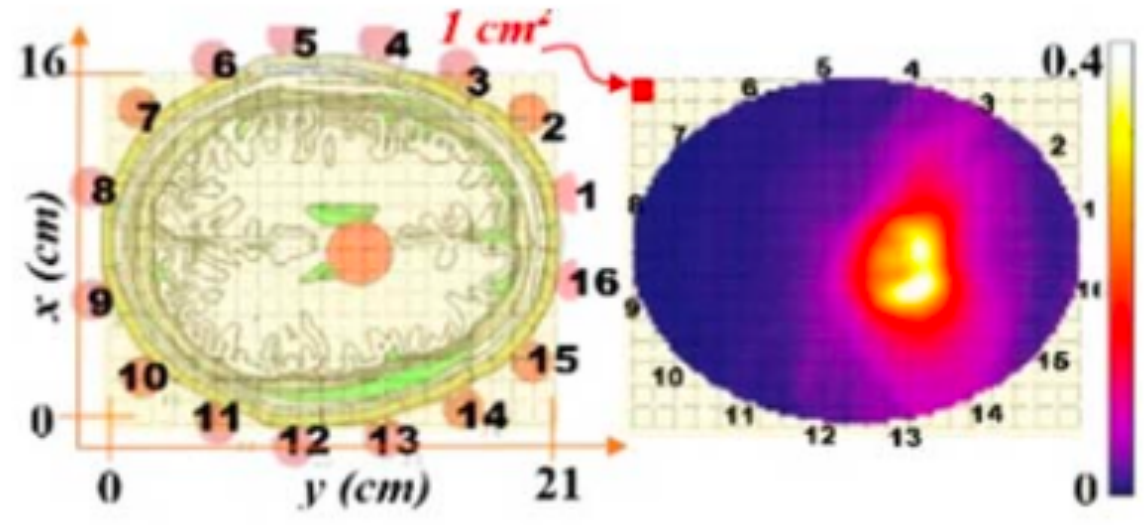

Reconstructed image from simulation.

Figure 49:

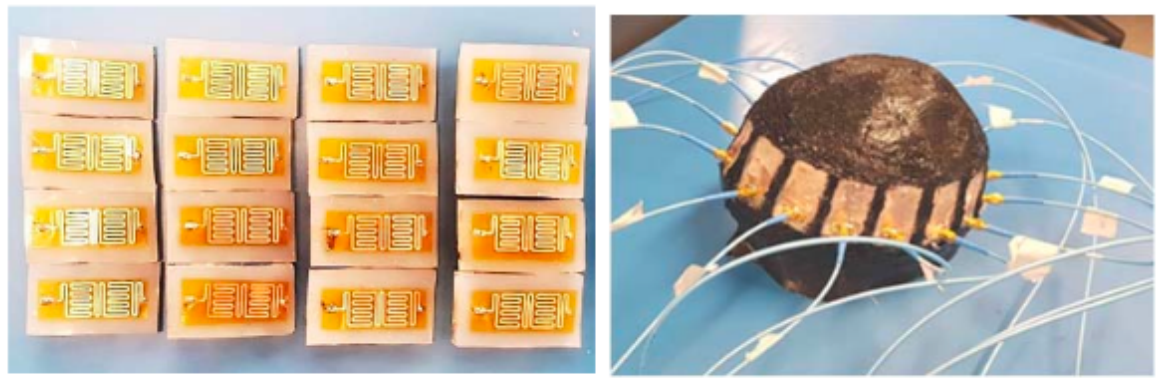

(a) (b)

16 antenna array elements of meander line shaped antenna are aligned in a head phantom model.

Figure 50:

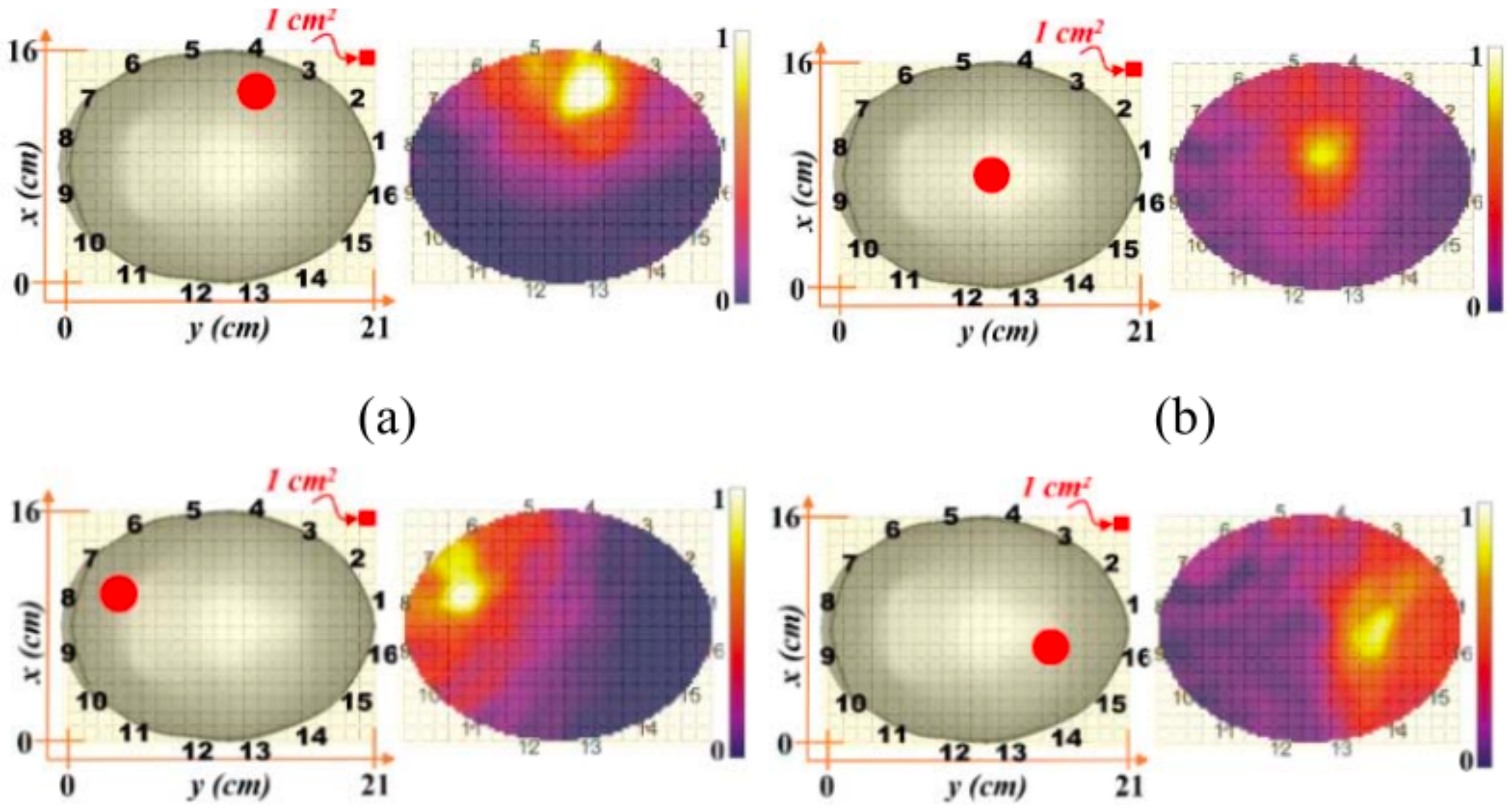

The image is reconstructed using DMAS reconstructing beamforming algorithm.

The good part of the proposed antenna is that the antenna include more compact width and lesser cutoff frequency(0.45GHz). They have used 16 arrays but 22 element arrays can be implemented due to smaller width of the antenna.

In the fourth paper we have reviewed A UWB Antenna for Microwave Brain Imaging by Adhitya Satria Pratama et.al. This paper is written in Indonesia, the targeted audience is in Indonesia where the population is very high and usage of Computed tomography method, position emission tomography magnetic resonance imaging is very costly for the general mass of population. The proposed antenna gives high frequency response, in the shape of dipole like structure fed by a coplanar waveguide. The antenna is simulated using CST microwave suite 2014 and there is a head phantom designed to measure the presence of brain tumour at a frequency of 5.8GHz. The proposed design in this paper UWB printed dipole shaped antenna is proposed by using conventional coplanar waveguide(CPW) feeding for frequency range from 3.1-10GHz for imaging.

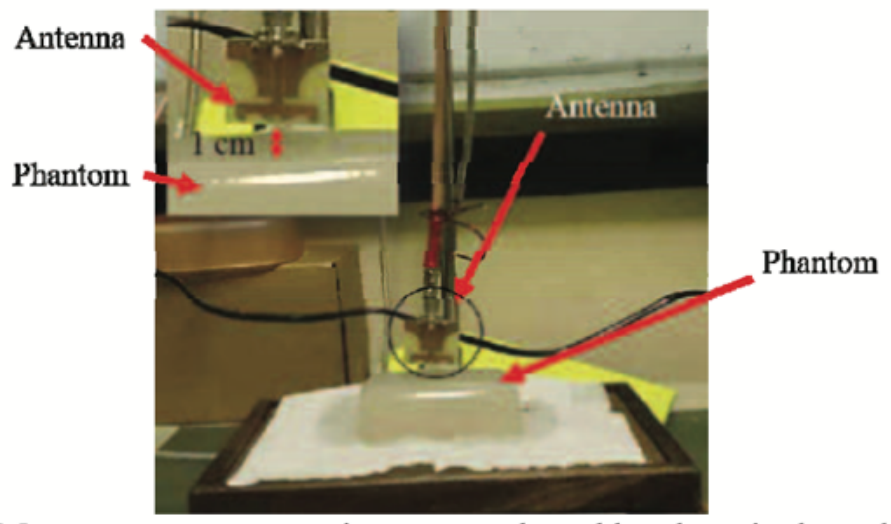

Fig 3. Measurement set-up using an agar-based head-equivalent phantom

In this the antenna is kept 5mm distance from the head phantom.

Figure 51:

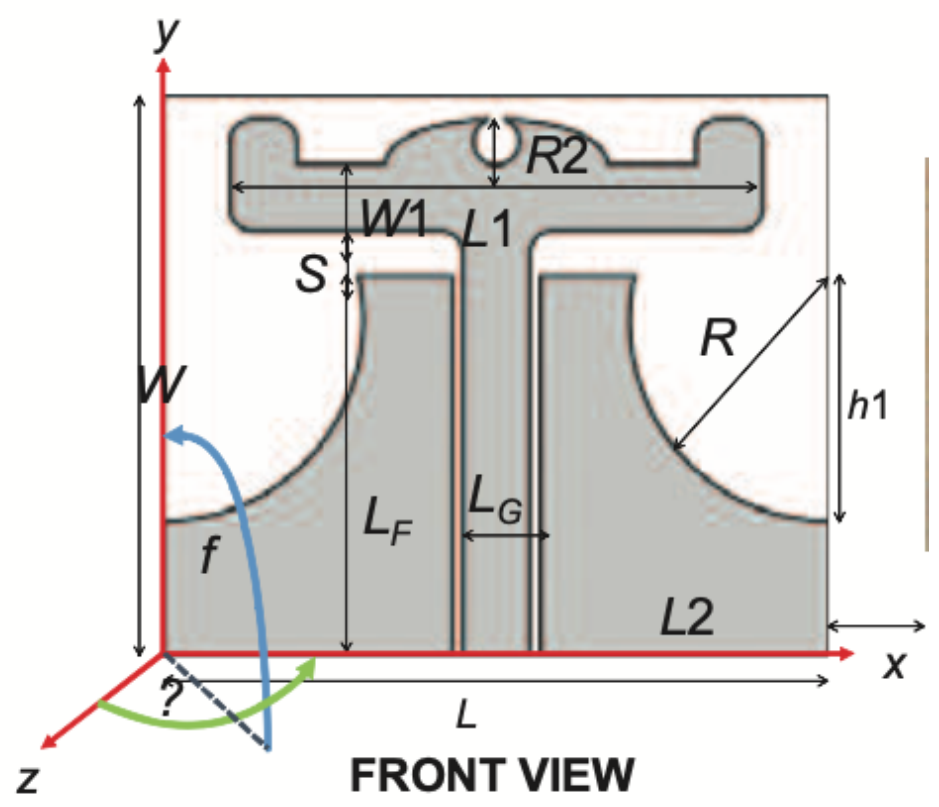

This is the dipole fed like antenna. Fabricated on a low cost FR-4 substrate with 1.6mm thickness. The size of the current proposed antenna is 25X30mm$^2$ which is comparatively small. The electrical conductivity of the head phantom is 34.93 of relative permittivity and 5.5195 S/m of conductivity with 35mm thinkness of the head phantom.

Figure 52:

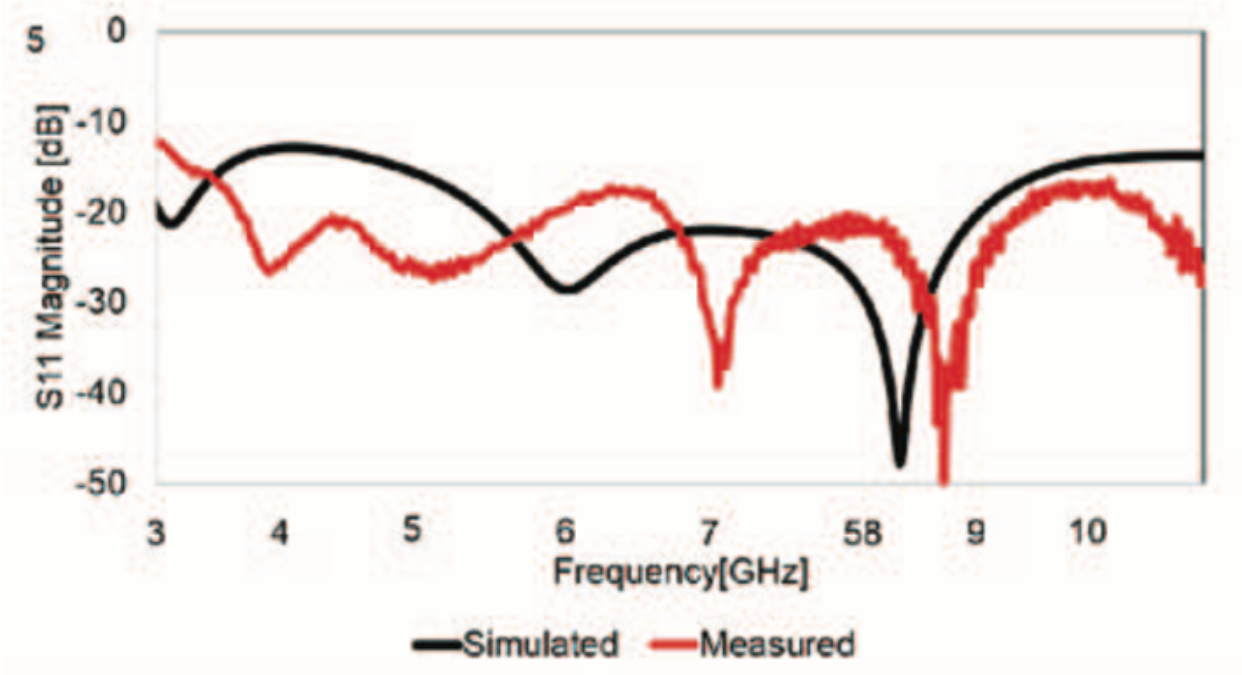

Fig 4. $S_{11}$ characteristics.

The $S_{11}$ parameters were obtained as follows. The simulated and measured values show coincidence and existence of the tumour in the brain at -50dB reflection coefficient.

In the fifth paper we have chosen High Performance Novel UWB Array Antenna for Brain Tumour Detection via Scattering Parameters in Microwave Imaging Simulation System by A H Ismail et.al. In this paper the S parameters are discussed for presence of tumour and absence of tumour in the human head. The antenna operated between 2.6GHz to 13.1GHz, the dimensions of the antenna is 80 X 45mm, in this dimension the maximum gain is calculated to be 12.12dB this is used to scan 9 distinct points on the human head. The S parameters are transformed using inverse fast Fourier transform and passed through filter for noise reduction. Brain tumour presence is indicated by the lower values of S parameters or reflection coefficient.

Figure 53:

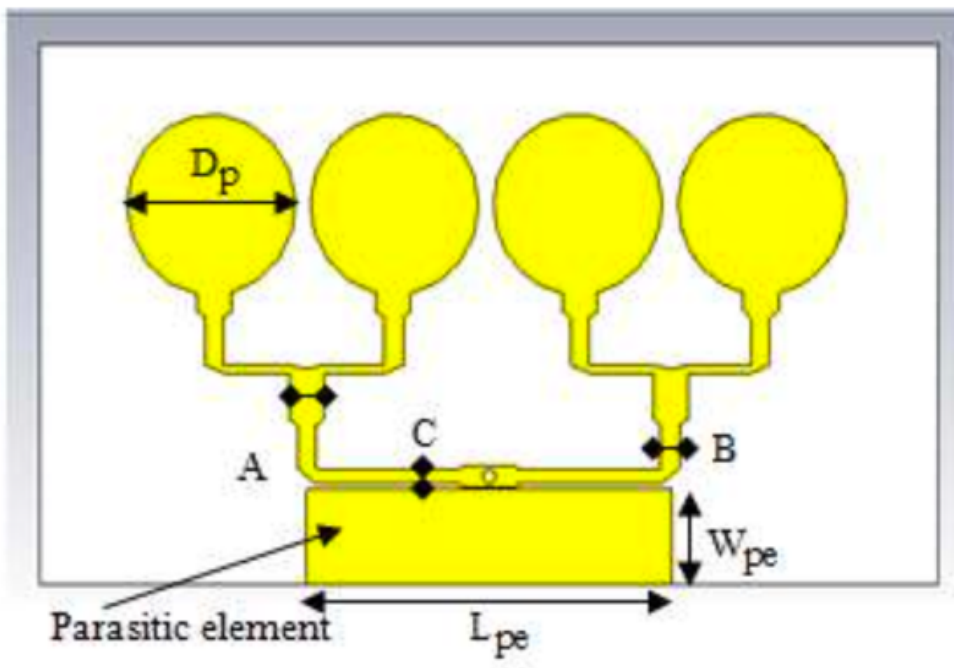

(a)

Copper radiating array with copper parasitic elements are used on the front and copper partial ground plane at the back side in association of 50 ohm SMA connector connected in concentric spaced circles as shown coaxially in figure.

There are four coaxial circular rings with diameter of 15mm with quarter wave transformer transmission line. In this design the small dimensions of 80mm X 45mm designed the antenna design. The parasitic element has width and length of 8mm and 32mm respectively, but in the ground plane, width and length were 18mm and 80mm respectively.

Figure 54:

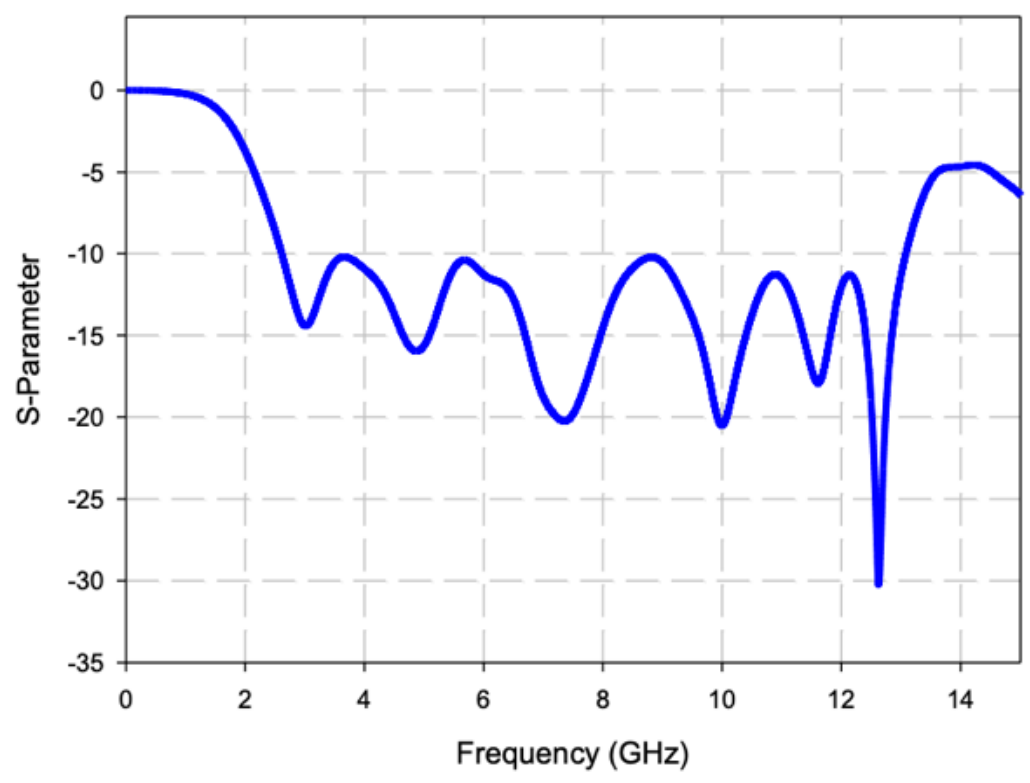

The S parameters are coming below -10dB and the received parameters are showing good obtained values at below -30dB.

There is an equal current distribution towards all four patches can be achieved by using the impedance matching. The scanning is first done with the tumour and then without the tumour. The nine positions of the human head is scanned.

Figure 54:

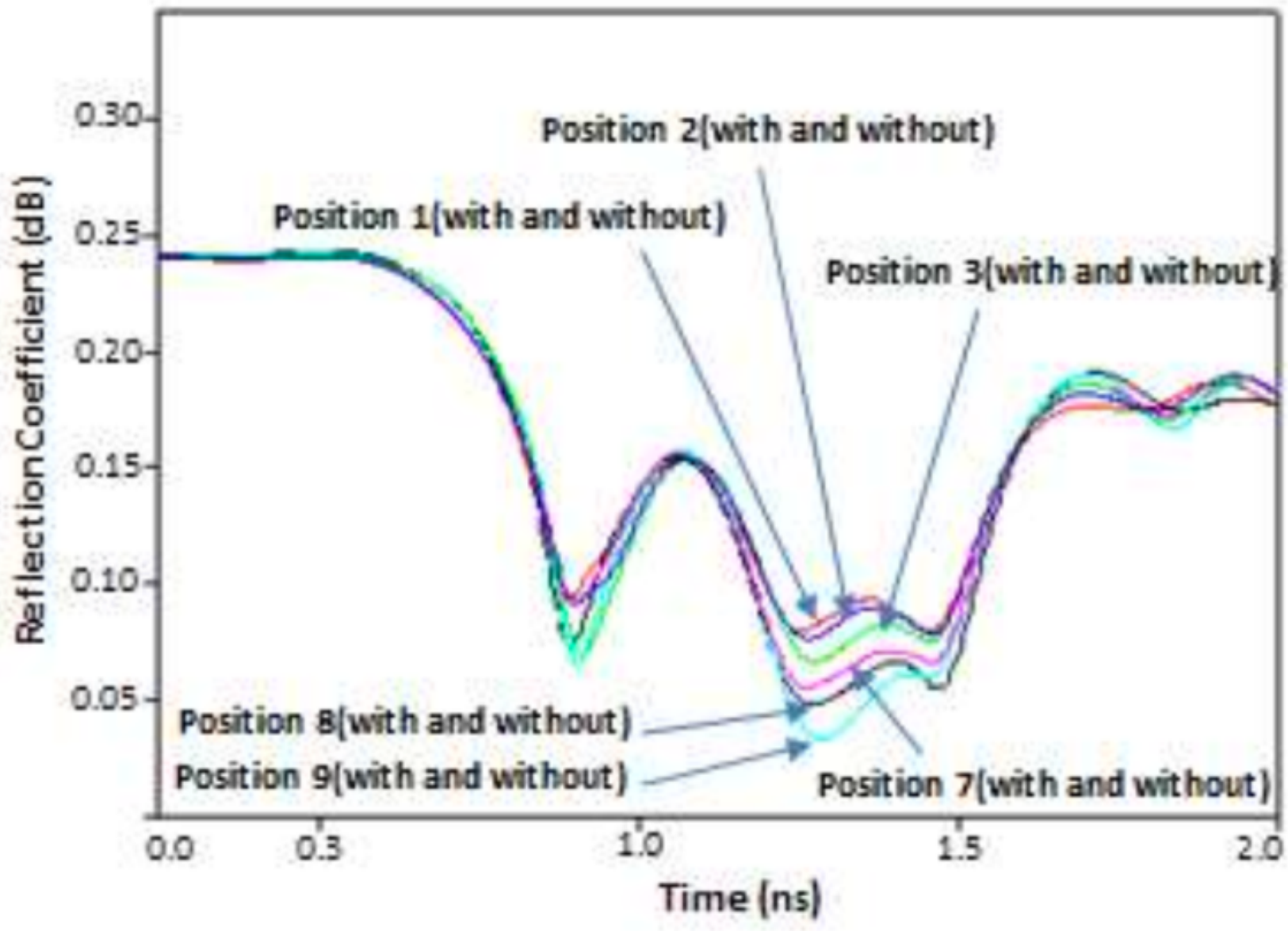

The different positions and the reflection coefficient values for nine different positions.

The light blue line indicates the presence of tumour at position 6. Position 5 and position 4 with and without tumour.

Figure 55:

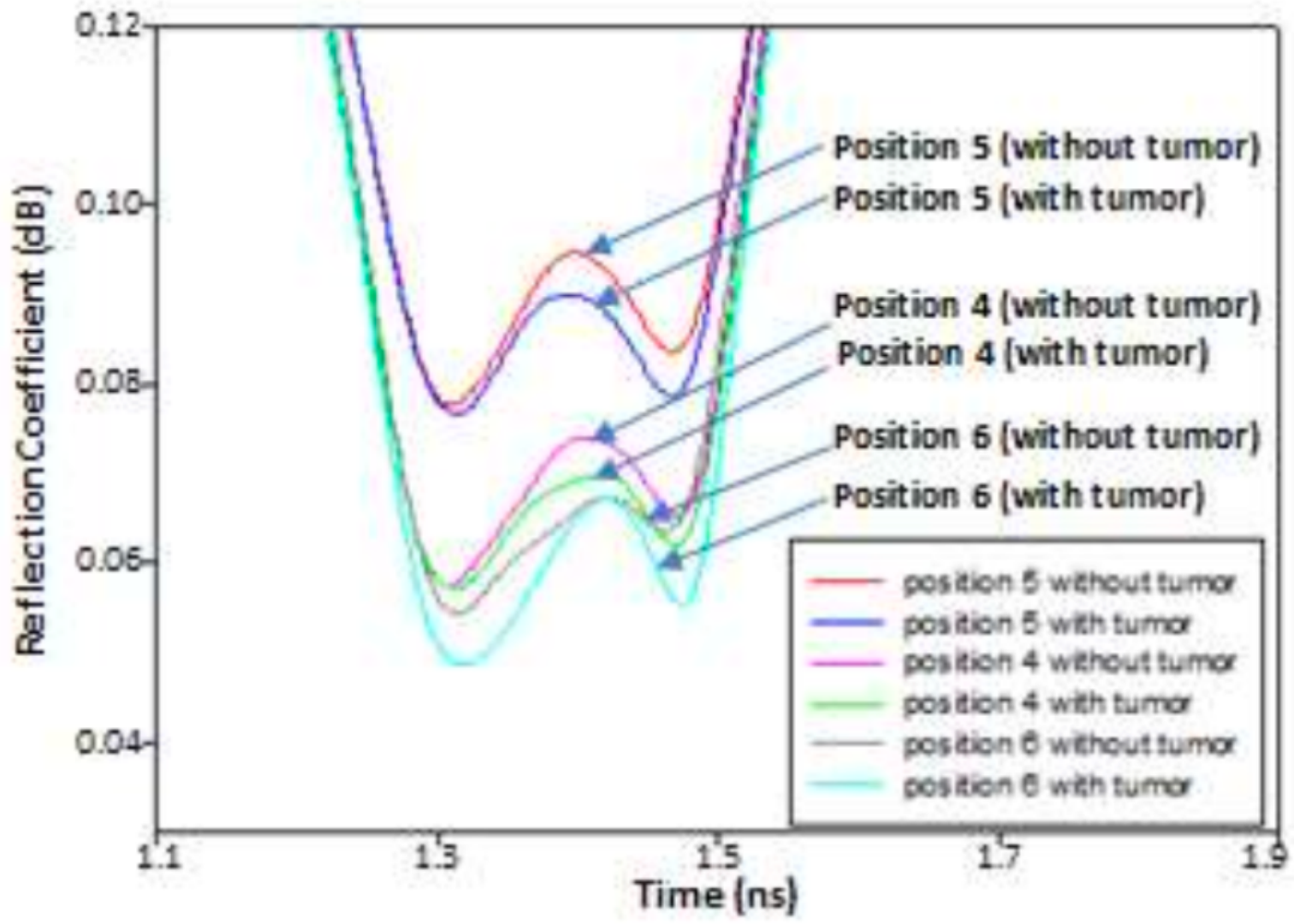

Fig. 11. Magnified axis view for positions 4, 5 and 6

In the sixth paper we have reviewed Microwave Imaging for Brain Tumour Detection Using an UWB Vivaldi Antenna Array by Haoyu Zhang et.al. In this paper an ultra wide band Vivaldi antenna is used for microwave brain imaging for cancer detection. The head phantom model is designed of four layers and a 5mm tumour is placed inside.

Figure 56:

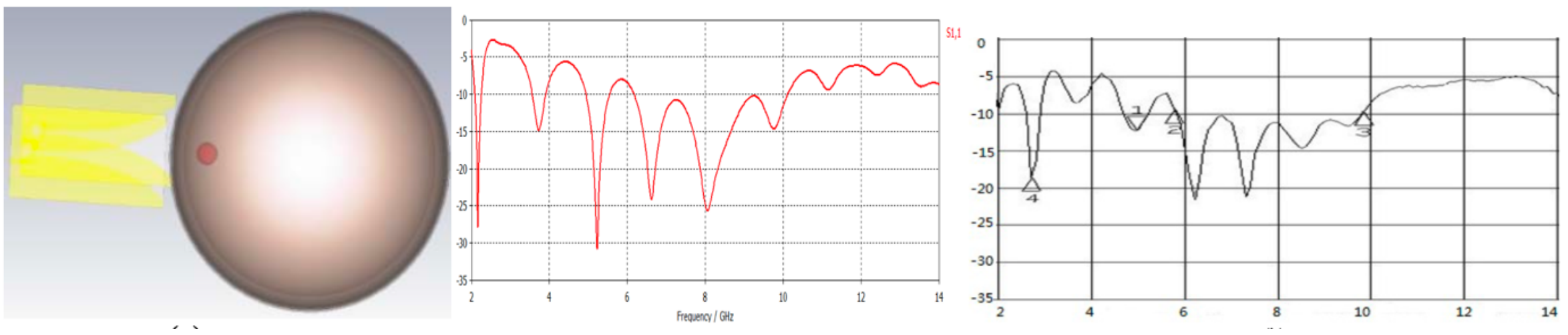

Figure showing head phantom with tumour, simulated $S_{11}$ parameters and measured $S_{11}$ parameters.

Figure 57:

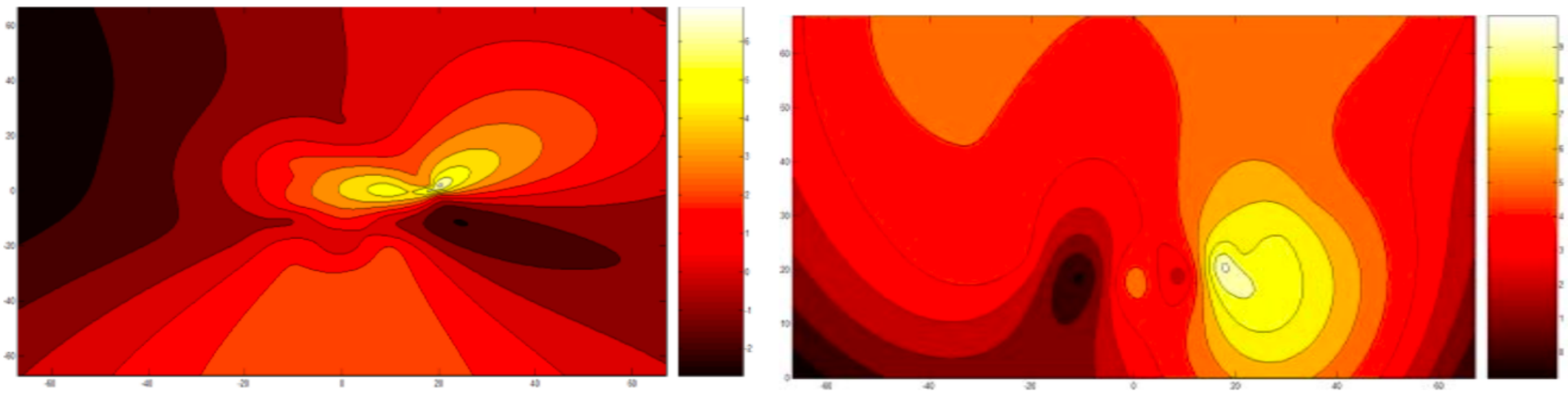

The reconstructed image. The tumour is distinct. Located at X=22, Y=21, Z=1. The actual tumour location is X=20, Y=20, Z=2. There is small error. The position of the tumour and detection of tumour is narrowly small.

In the seventh paper Microwave Imaging Using Frequency Domain Method for Brain Stroke Detection by A.M Abbosh et.al. The system consists of 24 antenna array tapered slot antennas. The antenna has 24X24X2 $mm^3$ dimensions, the operating frequency of the antenna is from 0.78 to 4GHz range which is analogous to 135% fractional bandwidth at centre 2.4GHz frequency. Antenna mutual coupling limitations is taken into account by keeping the antenna at 15 degrees separation.

Figure 58 and Figure 59:

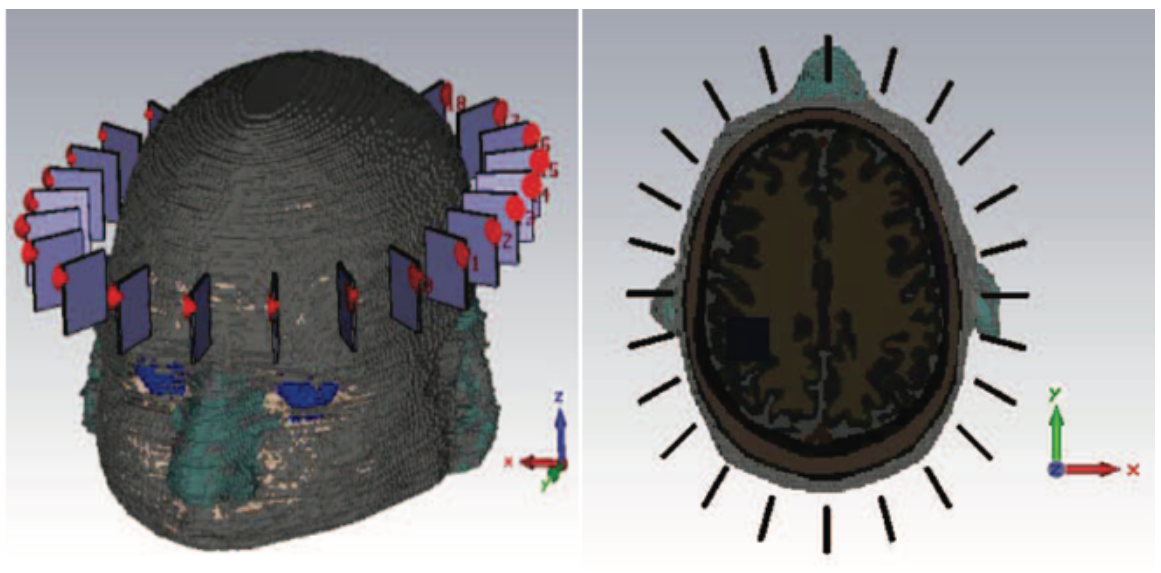

(a) (b)

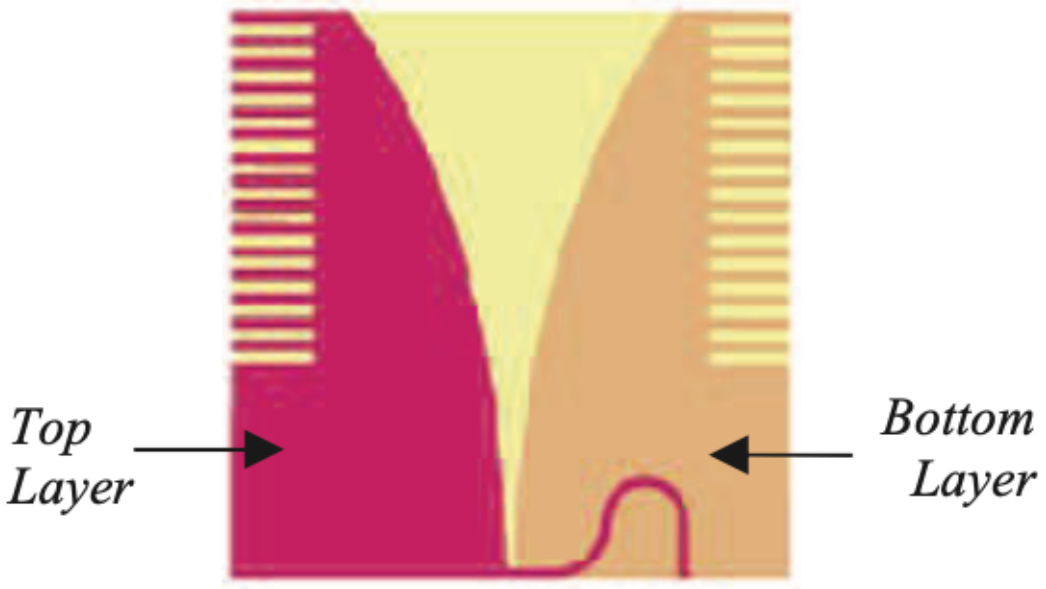

Figure 60 and Figure 61:

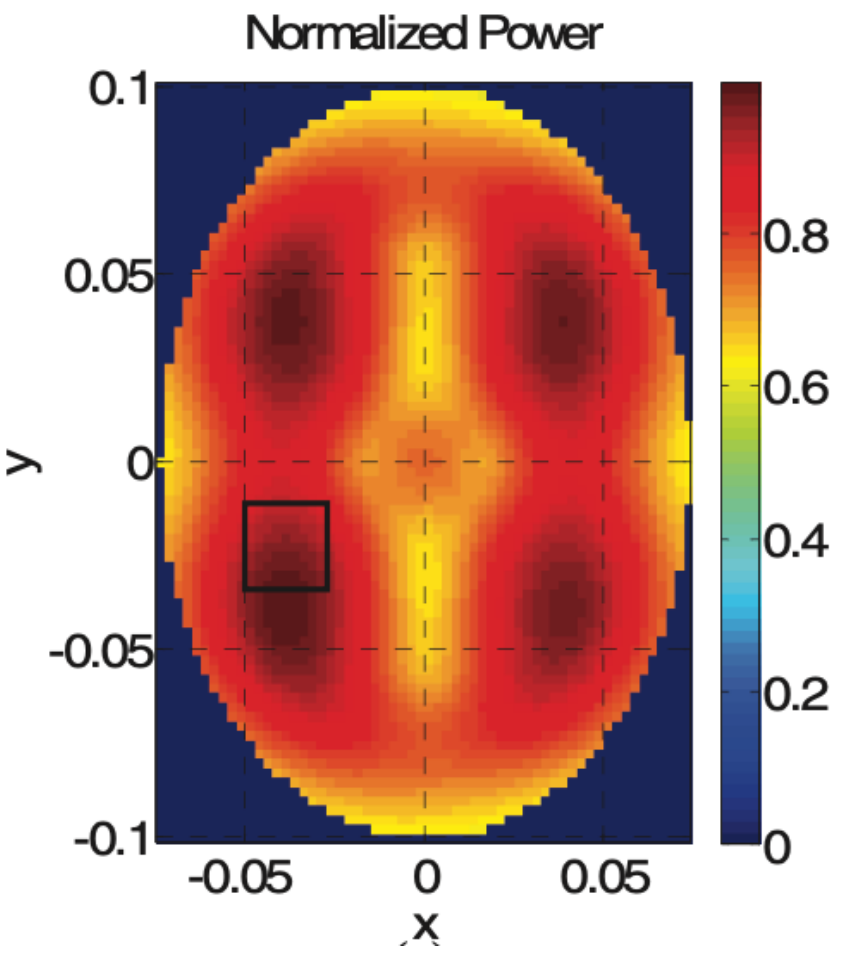

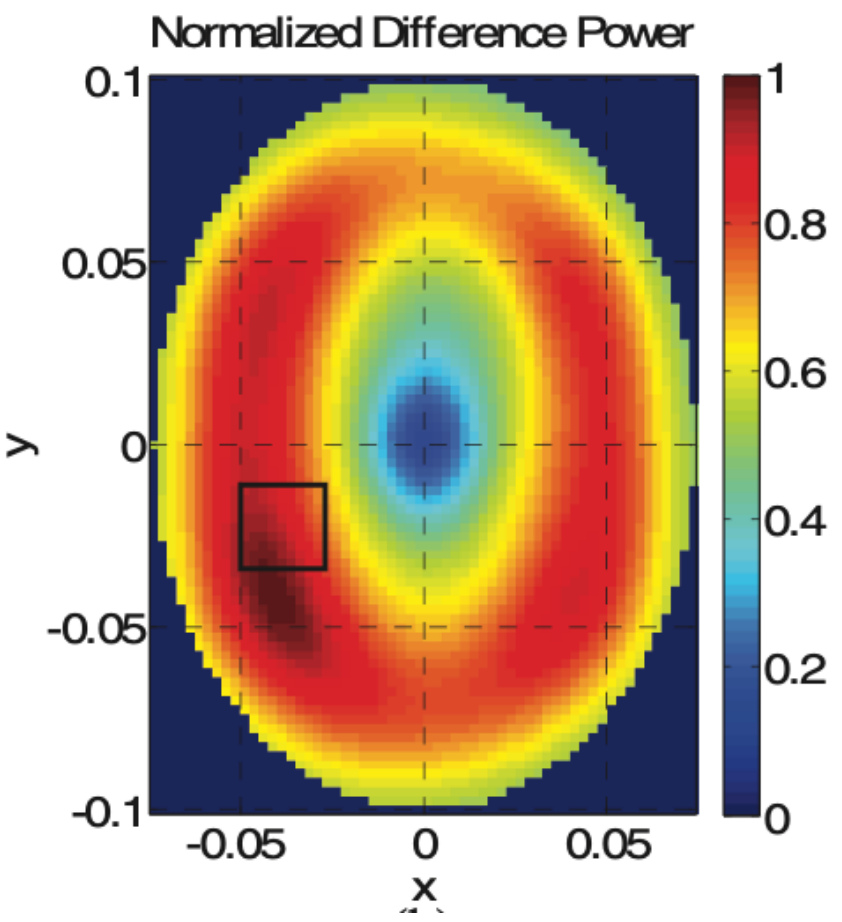

The normalized power and normalized difference power, the location of tumour from image is extracted. Mathieu function is utilized.

In the eighth paper titled Wavelet De-Noising Based Microwave Imaging for Brain Cancer Detection by Haoyu Zhang et.al they have shown image reconstruction technique using discrete wavelet transform(DWT) based signal processing to improve the noise performance of an ultra wide band antenna for brain tumour detection. The de-noised signals are used to create cross sectional analysis of the human head model with cancer.

Figure 62:

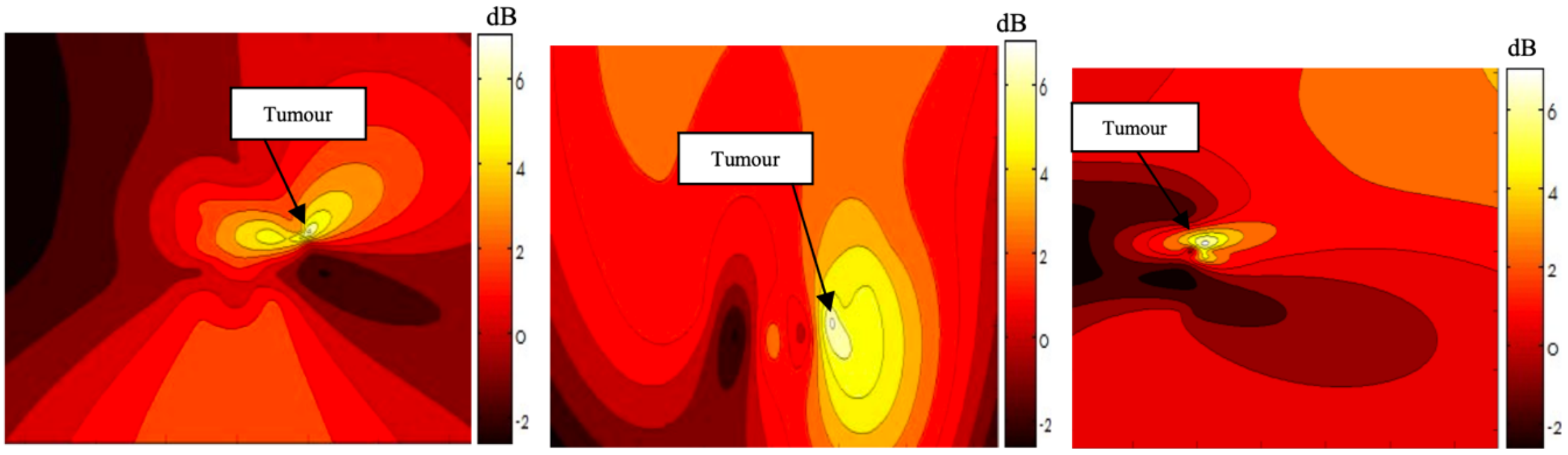

The tumour position a)x,y, no noise added b)y,z, no noise added c)y,z no noise added

Figure 63:

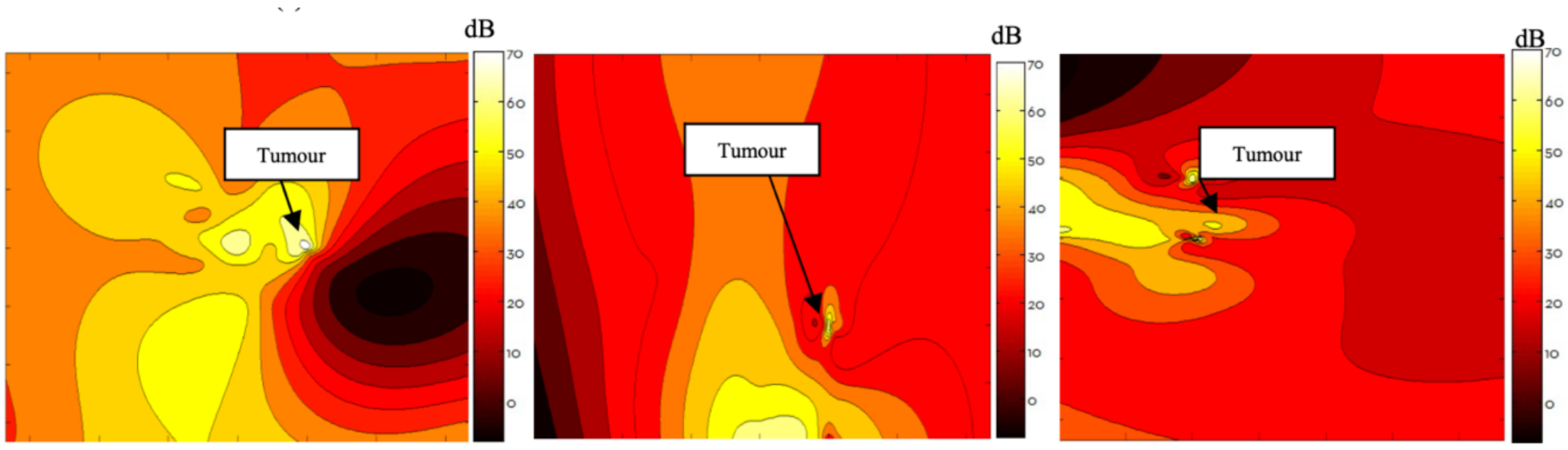

d)x,y, with noise added(SNR=60dB) e)x,z, noise added(SNR=60dB) f)y,z, noise added(SNR=60dB)

Figure 64:

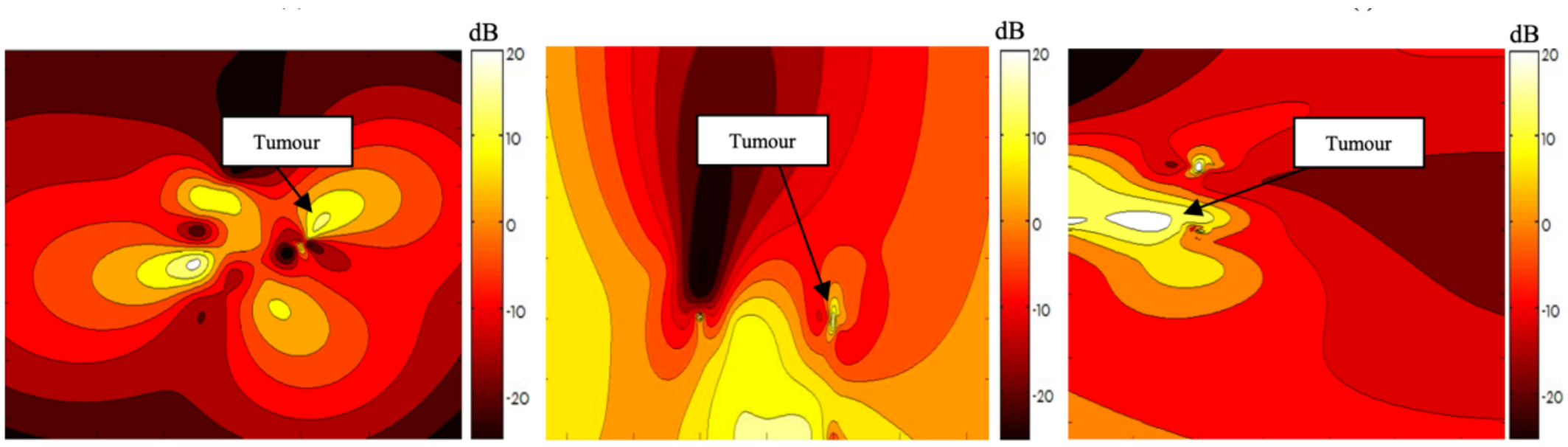

g)x,y, with noise added(SNR=45dB) h)x,z, noise added(SNR=45dB) i)y,z, noise added(SNR=45dB)

The retrieved images are as follows.

We have reviewed eight papers focusing primarily in antenna design for microwave brain imaging.

References:

[1] Samsuzzaman, M., Talukder, M.S., Alqahtani, A., Alharbi, A.G., Azim, R., Soliman, M.S. and Islam, M.T., 2023. Circular slotted patch with defected grounded monopole patch antenna for microwave-based head imaging applications. *Alexandria Engineering Journal*, *65*, pp.41-57.

[2] Hossain, A., Islam, M.T., Beng, G.K., Kashem, S.B.A., Soliman, M.S., Misran, N. and Chowdhury, M.E., 2022. Microwave brain imaging system to detect brain tumor using metamaterial loaded stacked antenna array. *Scientific Reports*, *12*(1), p.16478.

[3] Alqadami, A.S., Stancombe, A.E., Bialkowski, K.S. and Abbosh, A., 2020. Flexible meander-line antenna array for wearable electromagnetic head imaging. *IEEE Transactions on Antennas and Propagation*, *69*(7), pp.4206-4211.

[4] Pratama, A.S., Lubis, M.F.S., Zulkifli, F.Y. and Rahardjo, E.T., 2015, June. A UWB antenna for microwave brain imaging. In *2015 IEEE 4th Asia-Pacific Conference on Antennas and Propagation (APCAP)* (pp. 326-327). IEEE.

[5] Jamlos, M.A., Jamlos, M.F. and Ismail, A.H., 2015, April. High performance novel UWB array antenna for brain tumor detection via scattering parameters in microwave imaging simulation system. In *2015 9th European Conference on Antennas and Propagation (EuCAP)* (pp. 1-5). IEEE.

[6] Zhang, H., Flynn, B., Erdogan, A.T. and Arslan, T., 2012, November. Microwave imaging for brain tumour detection using an UWB Vivaldi Antenna array. In *2012 Loughborough Antennas & Propagation Conference (LAPC)* (pp. 1-4). IEEE.